\documentclass[applsci,article,accept,pdftex,moreauthors]{Definitions/mdpi} 
\usepackage{bm}
\usepackage{amsmath}
\usepackage{graphicx}
\usepackage{subcaption}
\usepackage{booktabs}
\usepackage{multirow}
\usepackage{makecell}
\usepackage{array}
\firstpage{1} 
\pubvolume{1}
\issuenum{1}
\articlenumber{0}
\pubyear{2024}
\copyrightyear{2024}
\datereceived{ } 
\daterevised{ } % Comment out if no revised date
\dateaccepted{ } 
\datepublished{ } 
\hreflink{https://doi.org/} % If needed use \linebreak

\Title{Comparative analysis of phase-field and intrinsic cohesive zone models for fracture simulations in multiphase materials with interfaces: Investigation of the influence of the microstructure on the fracture properties}
\TitleCitation{Title}

\Author{Rasoul Najafi Koopas$^{1,}$*, Shahed Rezaei$^{2}$, Natalie Rauter $^{1}$\orcidD{}, Richard Ostwald$^{3}$ and Rolf Lammering$^{1}$ }
 
\AuthorNames{Rasoul Najafi Koopas, Shahed Rezaei, Natalie Rauter, Richard Ostwald and Rolf Lammering}
\AuthorCitation{Najafi Koopas, R.; Rezaei, S.; Rauter, N.; Ostwald, R; Lammering, R.}
\address{%
$^{1}$ \quad Chair of Solid Mechanics, Institute of Mechanics, Helmut-Schmidt University/University of the Federal Armed Forces, Holstenhofweg  85, 22043 Hamburg, Germany; najafikr@hsu-hh.de (R.N.K.); natalie.rauter@hsu-hh.de (N.R.); rolf.lammering@hsu-hh.de (R.L.) \\
$^{2}$ \quad Access e.V., Aachen, Germany; s.rezaei@access-technology.de \\
$^{3}$ \quad Chair of Structural and Material Mechanics, Institute for Lightweight Design with Hybrid Systems, University of Paderborn, Pohlweg 47-49, 33098 Paderborn, Germany; richard.ostwald@uni-paderborn.de}

\corres{Correspondence: najafikr@hsu-hh.de}
 
\abstract{This study evaluates four widely used fracture simulation methods, comparing their computational expenses and implementation complexities within the Finite Element (FE) framework when employed on heterogeneous solids. Fracture methods considered encompass the intrinsic Cohesive Zone Model (CZM) using zero-thickness cohesive interface elements (CIEs), the Standard Phase-Field Fracture (SPFM) approach, the Cohesive Phase-Field fracture (CPFM) approach, and an innovative hybrid model. The hybrid approach combines the CPFM fracture method with the CZM, specifically applying the CZM within the interface zone. The finite element model studied is characterized by three specific phases: Inclusions, matrix, and the interface zone. This case study serves as a potential template for meso- or micro-level simulations involving a variety of composite materials. The thorough assessment of these modeling techniques indicates that the CPFM approach stands out as the most effective computational model provided that the thickness of the interface zone is not significantly smaller than that of the other phases. In materials like concrete, which contain interfaces within their microstructure, the interface thickness is notably small when compared to other phases. This leads to the hybrid model standing as the most authentic finite element model, utilizing CIEs within the interface to simulate interface debonding. A significant finding from this investigation is that within the CPFM method, for a specific interface thickness, convergence with the hybrid model can be observed. This suggests that the CPFM fracture method could serve as a unified fracture approach for multiphase materials when a specific interfacial thickness is used. In addition, this research provides valuable insights that can advance efforts to fine-tune material microstructures. An investigation of the influence of interfacial material properties, voids, and the spatial arrangement of inclusions shows a pronounced effect of these parameters on the fracture toughness of the material.}

\begin{document}

\section{Introduction}\label{sec1}
The non-linear behaviour of engineering materials results from their microstructural features. Fracture numerical simulations in small-scale level where these microstructural features are considered explicitly allow accurate predictions to be made without the need for complex material models. 

Two widely used numerical techniques for modelling crack initiation and propagation in small-scale brittle or semi-brittle materials are the phase-field fracture method (PFM) \cite{francfort1998revisiting, bourdin2000numerical} and the cohesive zone model (CZM) \cite{dugdale1960yielding, barenblatt1962mathematical, park2011cohesive}. Two PFM approaches are widely used to simulate crack initiation and propagation in brittle or semi-brittle materials namely as the Standard Phase-Field Fracture Model (SPFM) \cite{francfort1998revisiting, bourdin2000numerical} and the Cohesive Phase-Field Fracture Model (CPFM) \cite{lorentz2011convergence, lorentz2011gradient}. The main difference between SPFM and CPFM lies in the consideration of the length scale parameter, which determines the thickness of the diffused crack region. In SPFM, the length scale parameter is inherently dependent on the fracture toughness of the material. This dependence can lead to challenges in the accurate modeling of crack propagation, especially in materials with considerable heterogeneity \cite{miehe2010thermodynamically, conti2016phase, wu2017unified, GEELEN2019680}. In contrast, CPFM decouples the length scale parameter from the material properties, making it more suitable for small-scale simulations. In CPFM, crack propagation is controlled by a Traction-Separation Law embedded in the phase-field, which enables the modelling of cohesive cracking and debonding phenomena without predefined fracture zones \cite{verhoosel2013phase}. This makes CPFM particularly advantageous for the simulation of fractures in composite materials with interfaces.  

CZM is widely used to simulate fracture in brittle and semi-brittle materials, especially those with significant heterogeneity. CZM is divided into two categories: intrinsic and extrinsic approaches \cite{Ortiz1999, nguyen2014discontinuous}. The extrinsic method dynamically introduces CIEs during simulation, while the intrinsic method embeds CIEs in the material prior to simulation, making it effective for modelling weak interfaces. Interested readers are referred to  \cite{XU19941397, paggi2011nonlocal_I, paggi2011nonlocal_II, sane2018progressive, naghdinasab2018numerical, abbas20233d, wang2022microscale, guzel2023thermo, ozdemir2010thermo, dekker2021cohesive} for further practical applications of CZM. In this work, the intrinsic approach is used to define weak interfaces. Despite its effectiveness, the intrinsic CZM requires predefined fracture zones, which can be challenging for multiphase microstructures due to the complexity of implementation and computational cost.

\begin{figure}[t!]
\center
    \includegraphics[width=0.6\textwidth]{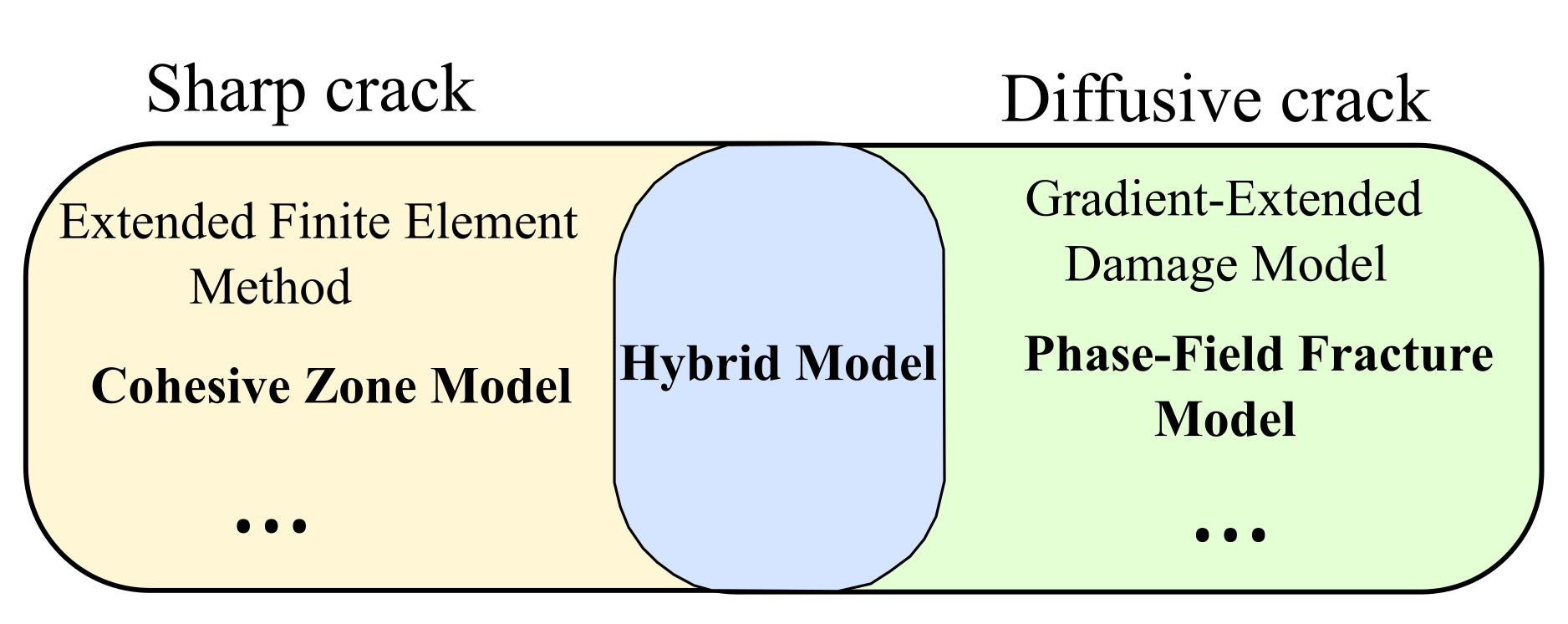}
    \caption{Application of finite element methods for simulating fracture in heterogeneous materials: The hybrid fracture approach integrates CZM for interface debonding and PFM for matrix cracking.}
    \label{introduction_pic}
\end{figure}

Recently, PFM has gained popularity in the simulation of crack initiation and propagation in multiphase materials with a interface phase. These simulations consider both matrix cracking and interface debonding. Based on how the interface is treated, two main approaches can be identified. In the first, the interface is approximated with a finite width and its failure is modelled using an auxiliary phase-field parameter \cite{yoshioka2021variational, pillai2023combined, hansen2019phase, nguyen2016phase, zhang2019modelling, li2020extension, li2020modeling, li2022phase, tarafder2020finite, dhaladhuli2022interaction, nguyen2022multi, zhou2022interface, yin2023modeling, bian2024unified, kumar2024modeling, nguyen2019phase}. Here, the interface fracture energy is derived from CZM or mathematical approximations taking into account other phases. In the second approach, the interface is assumed to have zero thickness and CIEs are used to simulate debonding \cite{paggi2017revisiting, zhang2020modelling, tan2021phase, hu2023phase, guillen2019micromechanical, guillen2020situ, han2023simulation}. This method avoids approximating the interface width, but requires the fracture zone to be identified in advance. Figure~\ref{introduction_pic} gives an overview of these fracture approaches.

Small-scale simulations face two main challenges: they are computationally expensive, especially for multiphase materials, and the simulation of crack initiation and propagation adds complexity that further increases the computational burden. This challenge is particularly significant in the FE2 approach, where performing small-scale simulations at each integration point in macroscale models is very resource intensive. Given these obstacles, it is crucial to choose a numerical approach that minimises complexity while reducing computational cost. The main contribution of this research is a comparative study of four numerical methods for simulating fracture in small-scale multiphase material microstructures in terms of implementation complexity and computational cost. The numerical models investigated include the CZM, SPFM, CPFM and a hybrid approach. The hybrid model combines CZM for interface debonding with CPFM for fracture in other phases. These methods are evaluated using finite element simulations of a single inclusion and a three-phase 2D benchmark model. The results show that optimising the positioning of inclusions, placing voids and adjusting interface properties can significantly influence crack propagation, resulting in increased energy dissipation and improved material toughness.
%#########################################################################
\section{Phase-Field Fracture Model for Brittle Fracture}
\begin{figure}[t]
\center
    \includegraphics[width=0.7\textwidth]{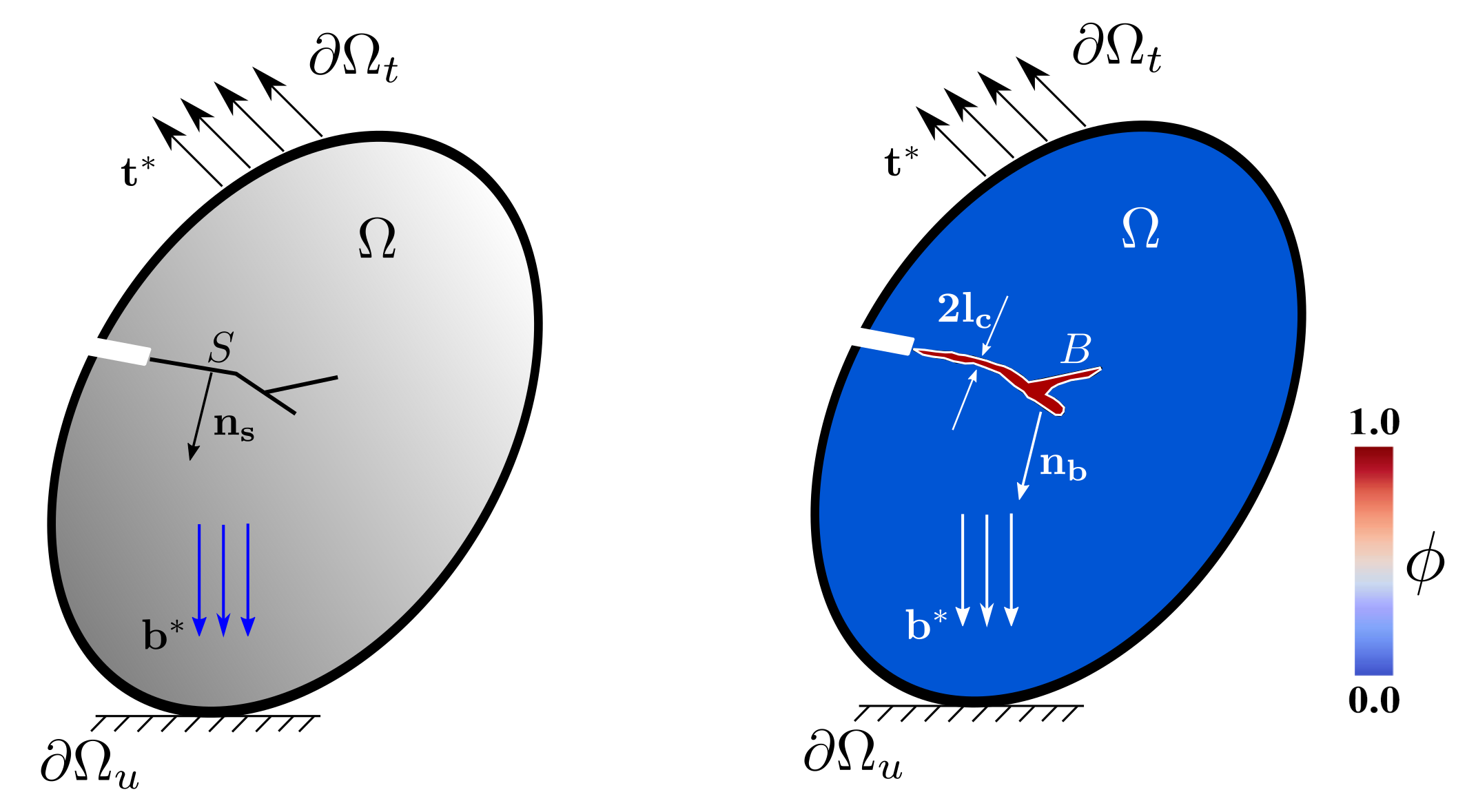}
    \caption{A solid undergoing fracture, regularized using a phase-field model: (a) Sharp crack representation (b) Phase-field regularized representation.}
    \label{formulation}
\end{figure}
As can be seen in Figure~\ref{formulation}, the phase-field fracture model approximates sharp cracks $S$ as diffuse bands whose thickness is defined by the length scale parameter \( l_c \). As \( l_c \) approaches zero, the diffuse crack reverts to a sharp crack \cite{braides1998approximation}. The phase-field variable \(\phi \in [0, 1]\) represents the degree of damage, with \(\phi = 0\) indicating no damage and \(\phi = 1\) a fully damaged state. As indicated in Equation \ref{irr_damage}, it is imperative that the time derivative of the phase-field damage parameter remains non-negative, ensuring the irreversibility of the induced damage, i.e. 
\begin{linenomath}
\begin{equation}  
   \dot{\phi}(\mathbf{x} ,t) \geq 0.
   \label{irr_damage}
\end{equation}
\end{linenomath}

Within the framework of the phase-field fracture method, the internal potential energy of the solid takes the form  \cite{bourdin2008variational}

\begin{linenomath}
\begin{equation}  
   \psi(\bm{\varepsilon}, \phi, \nabla_{\! \bm{x}} \phi) = \psi_{e}(\bm{\varepsilon}, \phi) + \psi_{c}(\phi, \nabla_{\!\bm{x}} \phi), 
\end{equation}
\end{linenomath}
where $\nabla_{\!\bm{x}} \bullet$ represents the spatial gradient of any scalar-valued field quantity $\bullet$, $\bm{\varepsilon}$ denotes the strain tensor pertaining to small deformations, $\psi_{e}$ denotes the stored strain energy, and $\psi_{c}$ represents the fracture surface energy. The determination of the stored strain energy is achieved through the relation

\begin{linenomath}
\begin{equation}  
    \psi_{e}(\bm{\varepsilon},\phi) = \omega(\phi)\,\psi_{0}(\bm{\varepsilon}),
    \label{elastic_energy}
\end{equation}
\end{linenomath}
where $\omega(\phi)$ represents the damage degradation function, responsible for reducing the stiffness of the undamaged material. The damage degradation function must adhere to the conditions
\begin{linenomath}
\begin{equation}  
    \omega(\phi=0)=1, \qquad \omega(\phi=1)=0, \qquad \omega'(\phi=1)=0.
    \label{constrains}
\end{equation}
\end{linenomath}

To mitigate the occurrence of cracking in regions subjected to compression, it is imperative to incorporate the damage degradation function in conjunction with the tension split of stored elastic energy such that

\begin{linenomath}
\begin{equation}
    \psi(\bm{\varepsilon}, \phi) = \omega(\phi)\,\psi_{0}^{+}(\bm{\varepsilon}) + \psi_{0}^{-}(\bm{\varepsilon}).
    \label{split}
\end{equation}
\end{linenomath}
where $\psi_{0}(\bm{\varepsilon}) = \frac{1}{2}\,\bm{\varepsilon}:\mathbb{E}_{0}:\bm{\varepsilon}$ represents the strain energy density function of the undamaged solid and $\mathbb{E}_{0}$ represents the fourth-order elastic stiffness tensor.

The fracture surface energy $\psi_{c}$ can be expressed as a volume integral via
\begin{linenomath}
\begin{equation}
    \psi_{c}(\phi, \nabla_{\!\bm{x}} \phi) = \int_\Omega G_{c}\gamma(\phi, \nabla_{\!\bm{x}} \phi) \,\mathrm{dA},
\end{equation}
\end{linenomath}
where $G_{c}$ is the critical energy release rate, and $\gamma(\phi, \nabla_{\!\bm{x}} \phi)$ is the fracture surface density function. The calculation of $\gamma(\phi, \nabla_{\!\bm{x}} \phi)$ is achieved using the relation

\begin{linenomath}
\begin{equation}
    \gamma(\phi, \nabla_{\!\bm{x}} \phi) =  \frac{1}{c_{0}}\left[\frac{1}{l_{c}}\alpha (\phi) + l_{c}\nabla_{\!\bm{x}} \phi \cdot \nabla_{\!\bm{x}} \phi\right],
    \label{surf_dens_func}
\end{equation}
\end{linenomath}
where $\alpha(\phi)$ is the crack geometric function, and is introduced to regularize $\psi_{c}(\phi, \nabla_{\!\bm{x}} \phi)$. 

The total potential energy of the solid domain is is expressed as

\begin{linenomath}
\begin{equation}
    P_{\mathrm{total}} = \int_{\Omega}^{} \omega(\phi)\,\psi_{0}(\bm{\varepsilon}) \,\mathrm{dV} + \int_{\Omega}^{} \frac{G_{c}}{c_{0}}\left[\frac{1}{l_{c}}\alpha (\phi) + l_{c}\,\nabla_{\!\bm{x}} \phi\cdot\nabla_{\!\bm{x}} \phi\right]  \,\mathrm{dV} -\int_\Omega \bm{b^*}\cdot \bm{u} \,\mathrm{dV}  - \int_\Omega \bm{t^*} \cdot \bm{u} \,\mathrm{dA}.
    \label{energy_functional}
\end{equation}
\end{linenomath}

The total energy functional represented by Equation \ref{energy_functional} is characterized as being quadratic and convex. Consequently, through the minimization of $P_{\text{total}}$, it is feasible to determine the displacement and phase-field fracture parameters denoted as $(\bm{u}, \phi)$ in a staggered manner. Within this model, crack propagation arises as a consequence of the interplay between the bulk and surface terms contained in $P_{\text{total}}$ according to
\begin{equation}
\label{energy_min}
    (\bm{u}, \phi) = \mathrm{arg}\{\mathrm{min}\, P_{\mathrm{total}}(\bm{u}, \phi)\} \quad \text{subjected to}\quad \dot\phi \geq 0 \quad \text{,}\quad \phi \in [0,1]. 
\end{equation}

Through implementation of the first variation of the total energy functional and the application of the divergence theorem, the strong form of the displacement field of the phase-field fracture model takes the form
\begin{linenomath}
\begin{equation}
  \begin{cases}
    \nabla \cdot \bm{\sigma} + \bm{b^*} = 0, \\
    \bm{u} = \bm{u}_{0} \quad \text{on } \partial\Omega_{u}, \\
    \bm{\sigma} \cdot \bm{n} = \bm{t} \quad \text{on } \partial \Omega_{t}.
  \end{cases}
  \label{strong1}
\end{equation}
\end{linenomath}

For the phase-field variable reads

\begin{linenomath}
\begin{equation}
  \begin{cases}
    \displaystyle Y-G_{c}\left[\frac{\partial\gamma(\phi, \nabla_{\!\bm{x}} \phi) }{\partial\phi}-\nabla_{\!\bm{x}} \cdot \left(\frac{\partial\gamma(\phi, \nabla_{\!\bm{x}} \phi)}{\partial\nabla_{\!\bm{x}}\phi}\right)\right] =0 & \text{if} \quad \dot\phi > 0, \\[5ex]
    \displaystyle Y-G_{c}\left[\frac{\partial\gamma(\phi, \nabla_{\!\bm{x}} \phi) }{\partial\phi}-\nabla_{\!\bm{x}} \cdot \left(\frac{\partial\gamma(\phi, \nabla_{\!\bm{x}} \phi)}{\partial\nabla_{\!\bm{x}}\phi}\right)\right] < 0 & \text{if} \quad \dot\phi = 0,\\
  \end{cases}
  \label{strong2}
\end{equation}
\end{linenomath}
where $Y$ is called energetic crack driving force and is determined from

\begin{linenomath}
\begin{equation}
    Y = -\frac{\partial\psi}{\partial\phi}=-\omega{'}(\phi)\frac{\partial\psi}{\partial\omega(\phi)}.
\end{equation}
\end{linenomath}

The difference between the SPFM and CPFM fracture approaches is determined based on the choice of three important variables, namely the crack geometric function, damage degradation function, and the history term variables. The corresponding formulations of CPFM and SPFM are elaborated in Appendix~\ref{Appendix_A}. Table~\ref{comparison} summarizes the corresponding phase-field fracture formulation for these two fracture approaches.
\begin{table}[H]
\caption{Comparison between standard and cohesive phase-field fracture models.\label{comparison}}
	\begin{adjustwidth}{-\extralength}{0cm}
		\newcolumntype{C}{>{\centering\arraybackslash}X}
		\begin{tabularx}{\fulllength}{lCC}
			\toprule
			\textbf{Parameters}	& \textbf{Standard Phase-Field Fracture}	& \textbf{Cohesive Phase-Field Fracture} \\
			\midrule
			$\phi(x)$  &  $\begin{cases} \phi(x) = \left(\frac{|x|}{2l_{c}}-1\right)^2 & \text{AT1 model} \\ \phi(x) = \exp\left(\frac{-|x|}{l_{c}}\right) & \text{AT2 model} \\ \end{cases}$ & $\phi(x) = 1-\sin\left(\frac{|x|}{l_{c}}\right)$ \\[5ex]
			$\alpha(\phi)$  & $\begin{cases} \alpha(\phi) = \phi & \text{AT1 model} \\ \alpha(\phi) = \phi^2 & \text{AT2 model} \\ \end{cases}$ & $\alpha(\phi) = 2\phi - \phi^2$ \\[5ex]
			$\omega(\phi)$ & $\omega(\phi) = (1-\phi)^2+k$ & $\omega(\phi) = \frac{(1-\phi)^2}{(1-\phi)^2+a_{1}\phi+a_{1}a_{2}\phi^2+a_{1}a_{2}a_{3}\phi^3}$ \\[5ex]
			$c_{0}$ & $\begin{cases} c_{0} = \frac{8}{3} & \text{AT1 model} \\ c_{0} = 2 & \text{AT2 model} \\ \end{cases}$ & $c_{0}=\pi$ \\[5ex]
			$H(x,t)$  & $H(\bold{x},t)= \max\left\{\max\psi_{\mathrm{eq}}(\bm{\varepsilon}(x, t))\right\}$ & $\begin{cases} H(\bold{x},t) = \max\left\{\psi_{\mathrm{eq}}^0\ , \max\psi_{\mathrm{eq}}(\bm{\varepsilon}(x, t))\right\} \\ \psi_{\mathrm{eq}}^0 = \frac{1}{2E}\sigma_{u}^2 \end{cases}$ \\[5ex]
			\bottomrule
		\end{tabularx}
	\end{adjustwidth}
\end{table}

% Cohesive zone model equations and their formatting
\subsection{Cohesive Zone Model}
Originally proposed by Barenblatt \cite{barenblatt1959formation, barenblatt1962mathematical}, the cohesive zone model focuses on the material behavior within the fracture process zone (FPZ) near the crack tip. In this region, significant interatomic forces arise due to the proximity of crack surfaces, depending on their relative displacements.

Numerical implementation of CZM in finite element analysis can involve duplicating nodes at interfaces between adjacent elements, creating a new cohesive interface element (segment-to-segment approach), or using the node-to-segment method \cite{paggi2016node}. CIEs follow a Traction-Separation law (TSL) to simulate crack behavior. For further details on CZM in heterogeneous materials, see \cite{najafi2023two}.

\subsection{Constitutive Equations of Cohesive Interface Elements}
\begin{figure}[t]
\center
    \includegraphics[width=0.6\textwidth]{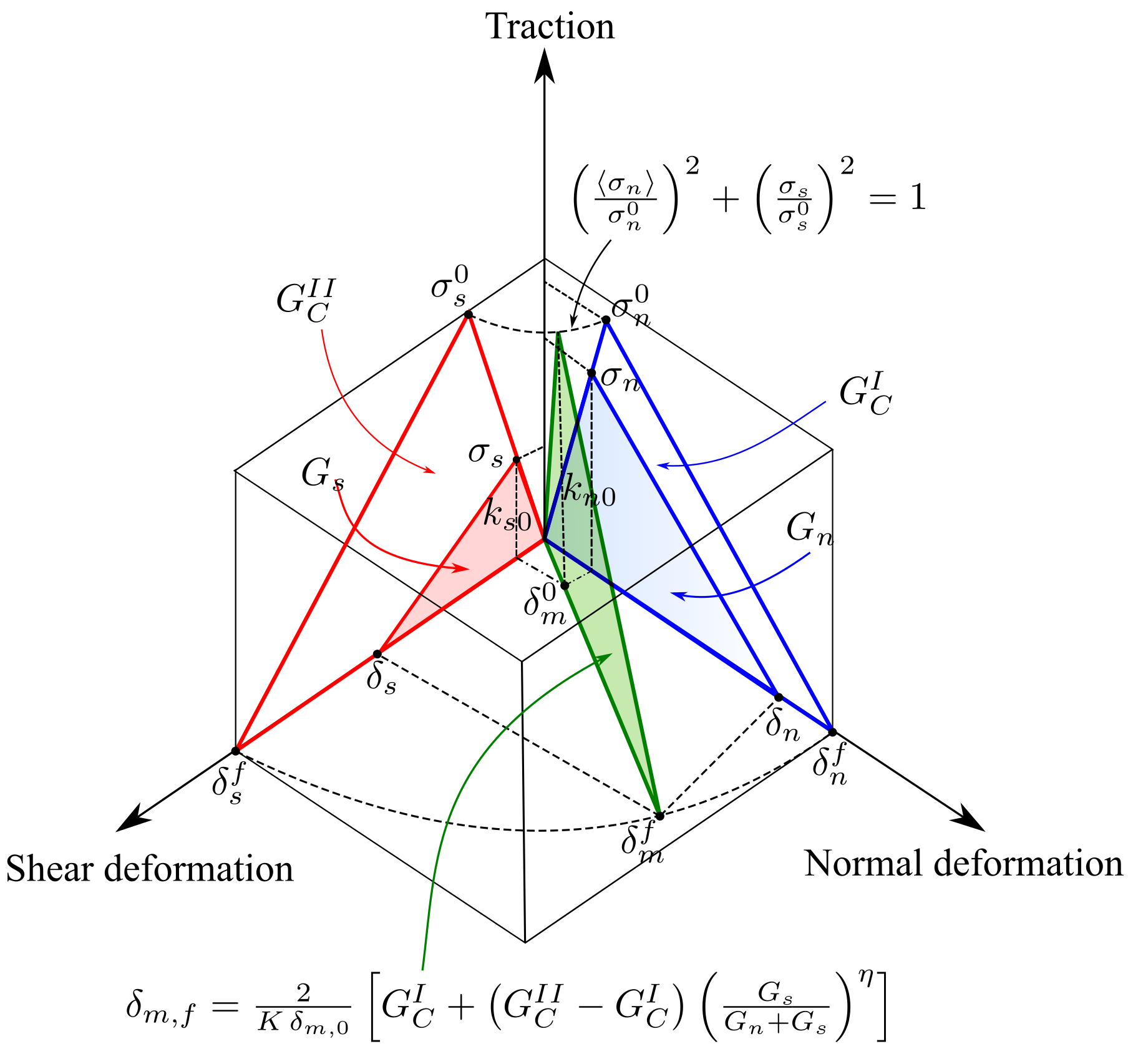}
    \caption{\label{validation}The quadratic nominal stress criterion and mixed-mode fracture criterion \cite{najafi2023two}.}
    \label{TSL_law}
\end{figure}

For multiaxial loading or mixed-mode fractures, material heterogeneities can induce mixed-mode crack propagation. Consequently, corresponding TSLs are used in finite element simulations \cite{xi2018meso}. As can be seen in Figure~\ref{TSL_law}, a bilinear TSL is characterised by penalty stiffness \(K\), threshold stress \(\sigma_{\mathrm{max}}\) and critical fracture energy \(G_{\mathrm{C}}\), which is adopted in this study. Fracture energy follows the form

\begin{linenomath}
\begin{equation}\label{fracEnergy}
    G_{C} = \int_{0}^{\delta_{f}} \sigma \, \mathrm{d}\delta \quad \text{for} \quad 0\le \delta\le\delta_{f}.
\end{equation}
\end{linenomath}
Before the initiation of damage, the normal and shear stress can be calculated as

\begin{linenomath}
\begin{equation}\label{elasticTraction_n}
\begin{cases}
\displaystyle \sigma_{n} = k_{n0}\left\langle \delta_{n} \right\rangle, & \text{(normal direction)} \\
\displaystyle \sigma_{t} = k_{s0} \, \delta_{s},  & \text{(shear direction)},  
\end{cases}
\end{equation}
\end{linenomath}
where \(k_{\mathrm{n0}}\) and \(k_{\mathrm{s0}}\) are penalty stiffnesses in the normal and shear directions. The Macaulay bracket \(\left\langle\bullet\right\rangle\) indicates that compressive deformation or stress states do not initiate damage in the model. A quadratic stress damage criterion is used for damage initiation. Damage is presumed to commence when the function involving nominal stress ratios attains a value of one

\begin{linenomath}
\begin{equation} \label{quadDamage}
   \left(\frac{\left\langle\ \sigma_{\mathrm{n}}\right\rangle }{\sigma_{\mathrm{n}}^0}\right)^2 + \left(\frac{\sigma_{\mathrm{s}}}{\sigma_{\mathrm{s}}^0}\right)^2 = 1,
\end{equation}
\end{linenomath}
where \(\sigma_{n}^0\) and \(\sigma_{s}^0\) are the tensile and shear cohesive strength, respectively. In FE simulations, when Equation (\ref{quadDamage}) is satisfied and applied displacement increases, both normal and shear stress diminish, signifying the onset of the softening region. A scalar damage variable 
\emph{D} is employed to quantify this stiffness degradation, encapsulating all damage mechanisms. \emph{D} monotonically evolves from 0 to 1, reaching 1 upon complete failure. The normal and shear stress in the softening region are calculated via  

\begin{linenomath}
\begin{equation}\label{elasticTraction}
\begin{cases}
\displaystyle \sigma_{n} = (1-D)\,k_{n0}\left\langle \delta_{n} \right\rangle & \text{(normal direction)}, \\
\displaystyle \sigma_{t} = (1-D)\,k_{s0} \, \delta_{s} & \text{(shear direction)}
\end{cases}
\end{equation}
\end{linenomath}

The evolution of the damage variable \emph{D} is computed using the expression proposed by \cite{camanho2002mixed}, i.e.

\begin{linenomath}
\begin{equation}  \label{DamageDeg}
   D = \frac{\delta_{\mathrm{m}}^f(\delta_{\mathrm{m}}^{\mathrm{max}}-\delta_{\mathrm{m}}^o)}{\delta_{\mathrm{m}}^{\mathrm{max}}(\delta_{\mathrm{m}}^{\mathrm{f}}-\delta_{\mathrm{m}}^o)}.
\end{equation}
\end{linenomath}
where \(\delta_{m}^0\) and \(\delta_{m}^f\) are the effective displacements at damage onset and complete failure, respectively, and \(\delta_{m}^{max}\) is the peak effective displacement during loading. The effective displacements in Equation (\ref{DamageDeg}) are computed using an expression suggested by \cite{camanho2002mixed} as

\begin{linenomath}
\begin{equation}  
   \delta_{m} =  \sqrt[]{\left\langle \delta_{n} \right\rangle^2 + \delta_{s}^2}.
\end{equation}
\end{linenomath}

The B-K fracture criterion \cite{benzeggagh1996measurement} for mixed-mode fractures (see Figure~\ref{TSL_law}) in composites and quasi-brittle materials calculates the effective displacement at failure as

\begin{linenomath}
\begin{equation} \label{BKCriteria}
   \delta_{m,f} = \frac{2}{K\delta_{m,0}}\Big[G_{C}^I+(G_{C}^{II}-G_{C}^I)\Big(\frac{G_{s}}{G_{n}+G_{s}}\Big)^\eta\Big],
\end{equation}
\end{linenomath}
where \(G_{C}^I\) and \(G_{C}^{II}\) are mode-I and mode-II fracture energies, and \(G_{n}\) and \(G_{s}\) are works from tractions and displacements in normal and shear directions, respectively. The parameter \(\eta\) is typically set to 1.2 for quasi-brittle materials \cite{camanho2002mixed}.
%#########################################################################
\section{Finite Element Modeling}\label{sec3}
\begin{figure}[t]
\center
    \includegraphics[width=0.7\textwidth]{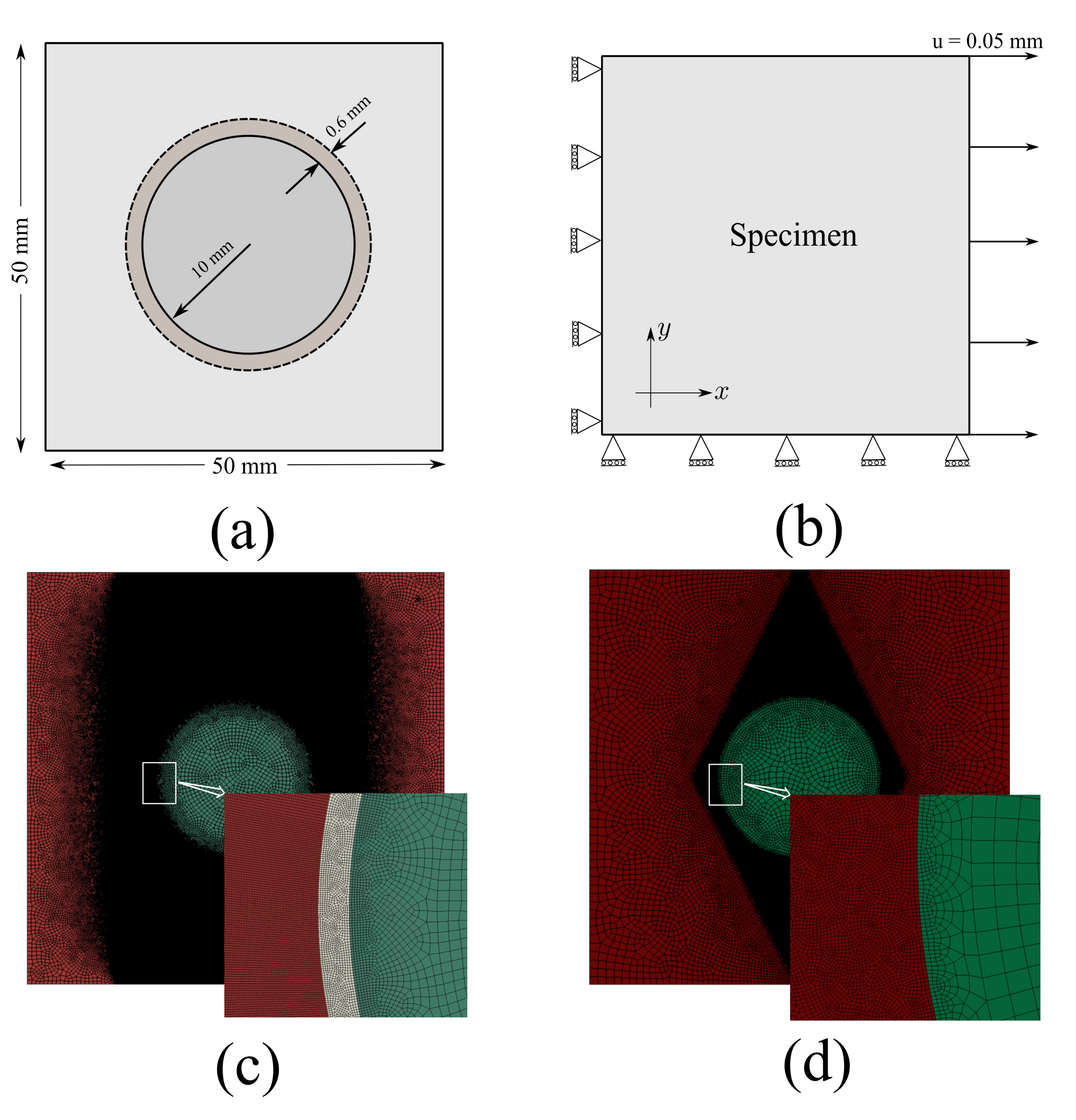}
    \caption{Geometrical features and boundary conditions of the 2D single inclusion benchmark model, with finite element discretization applied for phase-field fracture and cohesive zone model simulations. }
    \label{bc_gem}
\end{figure}

The multiphase 2D benchmark model in this study is discretized using the Abaqus/CAE Mesh module. The dimensions, applied boundary conditions and discretization strategies are shown in Figure~\ref{bc_gem}. An important aspect of the meshing process is ensuring continuity between the element surfaces, which must be maintained throughout the entire meshing process in CZM simulations. A four-node quadrilateral element (CPS4) by implementing free meshing technique is used for both the CZM and PFM simulations. For all the simulations, the 2D benchmark model is discretized with an average element length of 0.05 mm. For CZM simulations, once the model is meshed, in-house developed Python algorithms process the FE mesh file to generate zero-thickness cohesive interface elements across different phases. Further details on the CZM approach and the insertion of cohesive interface elements can be found in \citet{najafi2023two}. Similarly, for the phase-field fracture simulations, surface continuity is maintained across material phases, although in this case it is not strictly necessary. As the interface thickness is set to 0.6mm, meshing the model with an average element length of 0.05mm ensures that there are sufficient elements within the interface zone to accurately capture the gradient terms of the phase field damage. As shown in Figure~\ref{bc_gem}, all 2D specimens are subjected to a displacement of 0.05 mm in the $x$ direction.
%#########################################################################
Material properties for CZM and PFM simulations are given in Table \ref{czm_sim} and Table \ref{PFM_sim} respectively. Inclusions, known for their significantly higher mechanical strength, typically prevent crack propagation, a common observation in many engineering materials. As a result, the damage characteristics of inclusions are not explicitly defined in the CZM simulations. To reflect this behaviour in PFM simulations, the failure strength of inclusions is deliberately set much higher than that of the other material phases.
%#########################################################################
\begin{table}[t]
\caption{Material properties for CZM simulations. The properties of inclusion and matrix including Young's modulus, Poisson's ratio, and the 
\(\eta\) parameter, are derived from \cite{xiong2019meso}. The elastic stiffness of CIEs is referenced from \cite{wang2016computational}. The input parameters for the TSLs in relation to CIEs are obtained from \cite{najafi2023two}.}\label{czm_sim}
	\begin{adjustwidth}{-\extralength}{0cm}
		\newcolumntype{C}{>{\centering\arraybackslash}X}
		\begin{tabularx}{\fulllength}{lCCCCC}
			\toprule
			\textbf{Parameter} & \textbf{Inclusion} & \textbf{Matrix} & \makecell{\textbf{CIE-}\\\textbf{Interface}} & \makecell{\textbf{CIE-}\\\textbf{Matrix}} & \makecell{\textbf{CIE-}\\\textbf{Inclusion}} \\
			\midrule
			Elastic Modulus, \(E\) (GPa)  & 72 & 28 & - & - & - \\
			Poisson's ratio, \(\nu\) & 0.16 & 0.2 & - & - & - \\
			Elastic Stiffness, \(K_{n}\) (MPa/mm) & - & - & \(10^6\) & \(10^6\) & \(10^6\) \\
			Maximum normal stress, \(\sigma_{n}^0\) (MPa) & - & - & 2.4 & 4 & - \\
			Maximum shear stress, \(\sigma_{s}^0\) (MPa) & - & - & 10 & 30 & - \\
			Normal mode fracture energy, \(G_{I}\) (N/mm)  & - & - & 0.02 & 0.06 & - \\
			Shear mode fracture energy, \(G_{II}\) (N/mm)  & - & - & 0.4 & 1.2 & - \\
			B-K criterion material parameter, \(\eta\) & - & - & 1.2 & 1.2 & - \\
			\bottomrule
		\end{tabularx}
	\end{adjustwidth}
\end{table}
\begin{table}[t]
\caption{Material properties for phase-field simulations. The Young's modulus of the interface zone is obtained from \cite{xia2021mesoscopic}.}\label{PFM_sim}
	\begin{adjustwidth}{-\extralength}{0cm}
		\newcolumntype{C}{>{\centering\arraybackslash}X}
		\begin{tabularx}{\fulllength}{lCCCC}
			\toprule
			\textbf{Phase} & \makecell{\textbf{Young’s modulus} \\ \boldmath$E$ \textbf{(GPa)}} & \makecell{\textbf{Poisson’s ratio} \\ \boldmath$\nu$} & \makecell{\textbf{Fracture energy} \\ \boldmath$G_{c}$ \textbf{(N/mm)}} & \makecell{\textbf{Failure strength} \\ \boldmath$\sigma_{u}$ \textbf{(MPa)}} \\
			\midrule
			Inclusion  & 72 & 0.16 & 0.2 & 20 \\
			Matrix  & 28 & 0.2  & 0.06  & 4 \\
			Interface  & 21.9 & 0.2  & 0.02  & 2.4 \\
			\bottomrule
		\end{tabularx}
	\end{adjustwidth}
\end{table}

In this study, a staggered scheme is used for phase-field fracture simulations, where the phase-field and displacement fields are solved sequentially. The staggered scheme proposed by \citet{molnar20172d} is used to solve the system of nonlinear equations for PFM simulations. In addition, the CPFM approach is integrated into the ABAQUS UEL subroutine using the open source ABAQUS codes from \citet{molnar20172d}, with modifications applied to the UEL subroutine to convert the code from the AT2 standard phase-field fracture model to the CPFM fracture model. These modifications are based on the equations presented in Table \ref{comparison}. 

Similar to other studies, such as \cite{wang2016computational, esmaeeli2019two} the Abaqus Explicit solver is used for the CZM simulations. The reason for this choice is that the insertion of CIE between each two solid elements introduces a high degree of nonlinearity into the model, making it very difficult to use static, quasi-static or implicit solvers. However, in this study, to ensure that dynamic effects are not present in the CZM simulation, two strategies are implemented: first, the displacement is applied gradually, and second, the time increment is kept very small ( in this study is $2.99 \times 10^{-08}$). With these strategies, the kinetic energy in the CZM simulations remains around 2\% of the total energy of CZM simulation.

Moreover, reaction force-displacement curves are derived by averaging the reaction forces of the nodes along the left vertical edge, and stress-strain curves are obtained by dividing the reaction force by the element edge length of the benchmark model.
%#########################################################################
\section{Results}\label{sec4}
The main focus of this section is on assessing the numerical efficiency of the CZM, CPFM, SPFM and hybrid approaches. Afterward, the best fracture model is extended to more complex scenarios with multiple randomly distributed inclusions, while maintaining consistent boundary conditions and material properties. To further demonstrate the feasibility of applying the best fracture model in 3D, a three-dimensional benchmark model using the most efficient scheme is also included.
\begin{figure}[t]
\center
    \includegraphics[width=0.5\textwidth]{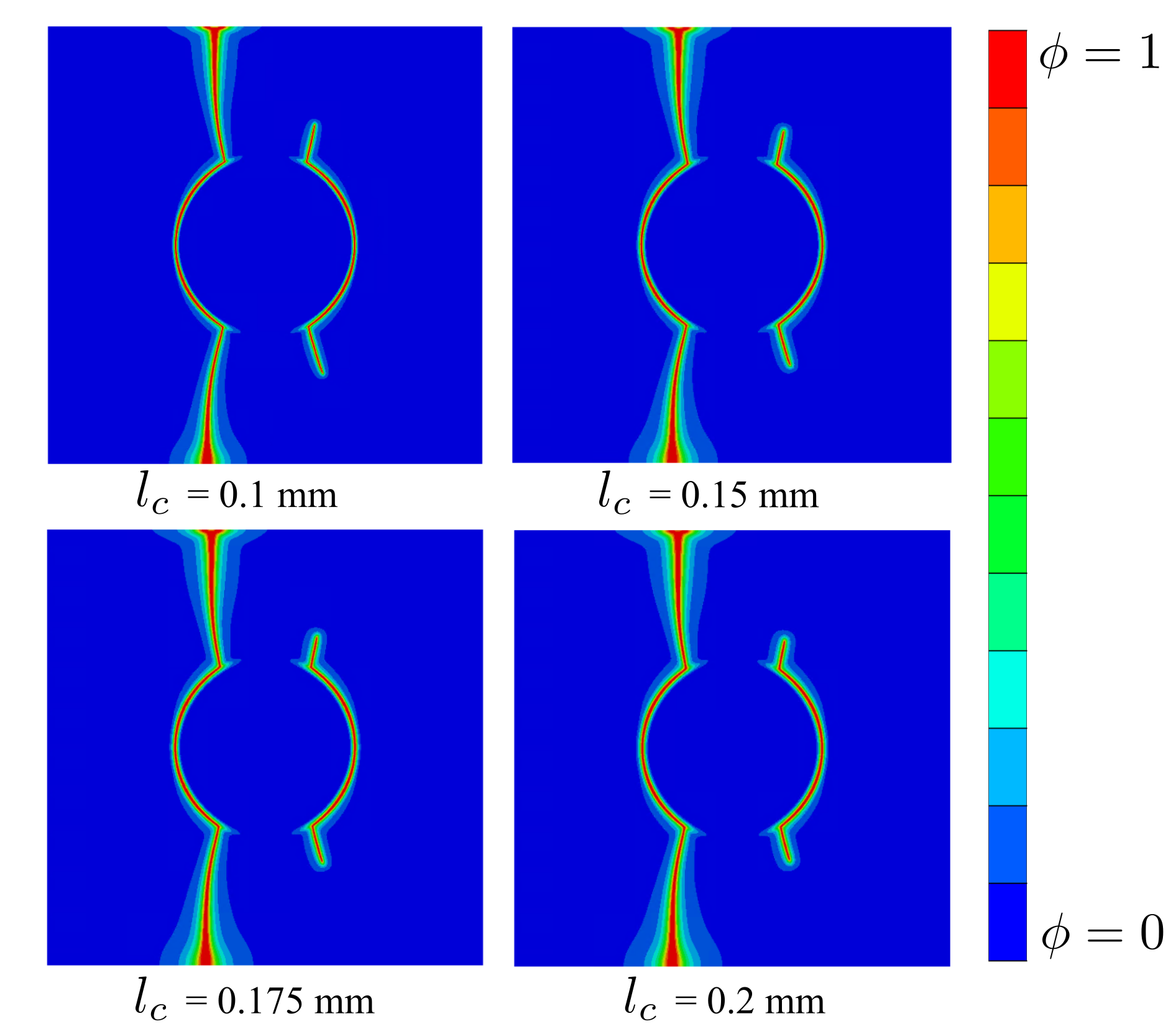}
    \caption{Fracture paths in the 2D single inclusion model have been simulated under tensile loading conditions. Here, \(\phi\) denotes the phase-field fracture parameter, with \(\phi = 0\) indicating no damage, and \(\phi = 1\) illustrating a complete fracture.}
    \label{spfm}
\end{figure}

\begin{figure}[H]
\center
    \includegraphics[width=0.6\textwidth]{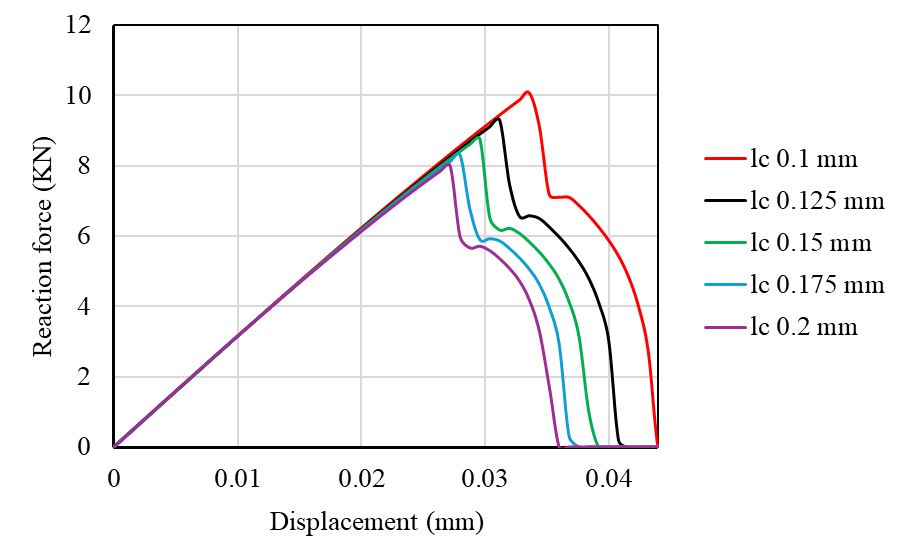}
    \caption{Sensitivity analysis of the parameter \(l_{c}\) conducted using the AT2 standard phase-field fracture model.}
    \label{lc_spfm}
\end{figure}

%#########################################################################
\subsection{$l_{c}$ sensitivity analysis}
In the phase-field fracture model, the fracture behaviour is governed by the minimisation of the total energy functional, which determines both the displacement \(\bm{u}\) and the phase-field parameter \(\phi\). Unlike CZM, the fracture path in phase-field fracture models is independent of the mesh configuration and the diffused fracture width is controlled by the length scale parameter \(l_{c}\). 
However, the limitations of the SPFM approach become apparent in small-scale simulations of multiphase materials. In the following sections, the effect of the parameter \(l_{c}\) in small-scale standard and cohesive phase-field fracture simulations is compared, and its limitations are analyzed.
%#########################################################################
\subsubsection{Standard phase-field fracture}
In this analysis, a sensitivity study of the length scale parameter \(l_{c}\) is performed using a 2D benchmark model within the AT2 standard phase-field fracture framework. All parameters are held constant while \(l_{c}\) is systematically varied between 0.1 mm and 0.2 mm. Figure~\ref{spfm} shows the simulated fracture paths for different \(l_{c}\) values, with a mesh size of 0.05 mm used in the finite element analysis. The reaction force-displacement curves, as shown in Figure~\ref{lc_spfm}, reveal two key observations consistent with \citet{nguyen2016choice}. Firstly, an increase in \(l_{c}\) leads to a reduction in the peak reaction force. Secondly, across different \(l_{c}\) values, a sharp decline in reaction force is observed after the peak, with only a small increase in displacement.

It is important to note that \(l_{c}\) cannot only be considered as a numerical parameter within the SPFM approach. The relationship between \(l_{c}\) and material strength has been clarified by simplified analytical solutions \cite{amor2009regularized, kuhn2014simulation, borden2012phase}, while experimental studies have determined \(l_{c}\) by measuring the critical stress in uniformly stretched specimens \cite{nguyen2016choice}. Using Irwin's equation, \(l_{c}\) can be calculated from material properties such as Young's modulus \(E\), fracture energy \(G_c\) and tensile strength \(\sigma\) for the SPFM simulations:

\begin{linenomath}
\begin{equation}  
   l_c = \frac{27 E G_c}{8 \sigma^2}
\end{equation}
\end{linenomath}

Considering Table \ref{PFM_sim}, the calculated length scale parameter for the inclusion phase is \(l_c = 0.243 \, \text{mm}\). Similarly, for the matrix phase the length scale parameter is \(l_c = 0.142 \, \text{mm}\) and for the interfacial phase it is \(l_c = 0.128 \, \text{mm}\). As these results suggest, the \(l_{c}\) values are not appropriate for small-scale simulations due to the size of the specimen. Using a larger \(l_{c}\) in small-scale simulations can introduce boundary effects and inaccuracies, particularly in the matrix and interfacial phases where the calculated \(l_{c}\) values are relatively small. This highlights the limitations of the SPFM approach for small-scale simulations, especially for heterogeneous multi-phase microstructures.
%#########################################################################
\subsubsection{Cohesive phase-field fracture}
\begin{figure}[t]
\center
    \includegraphics[width=0.5\textwidth]{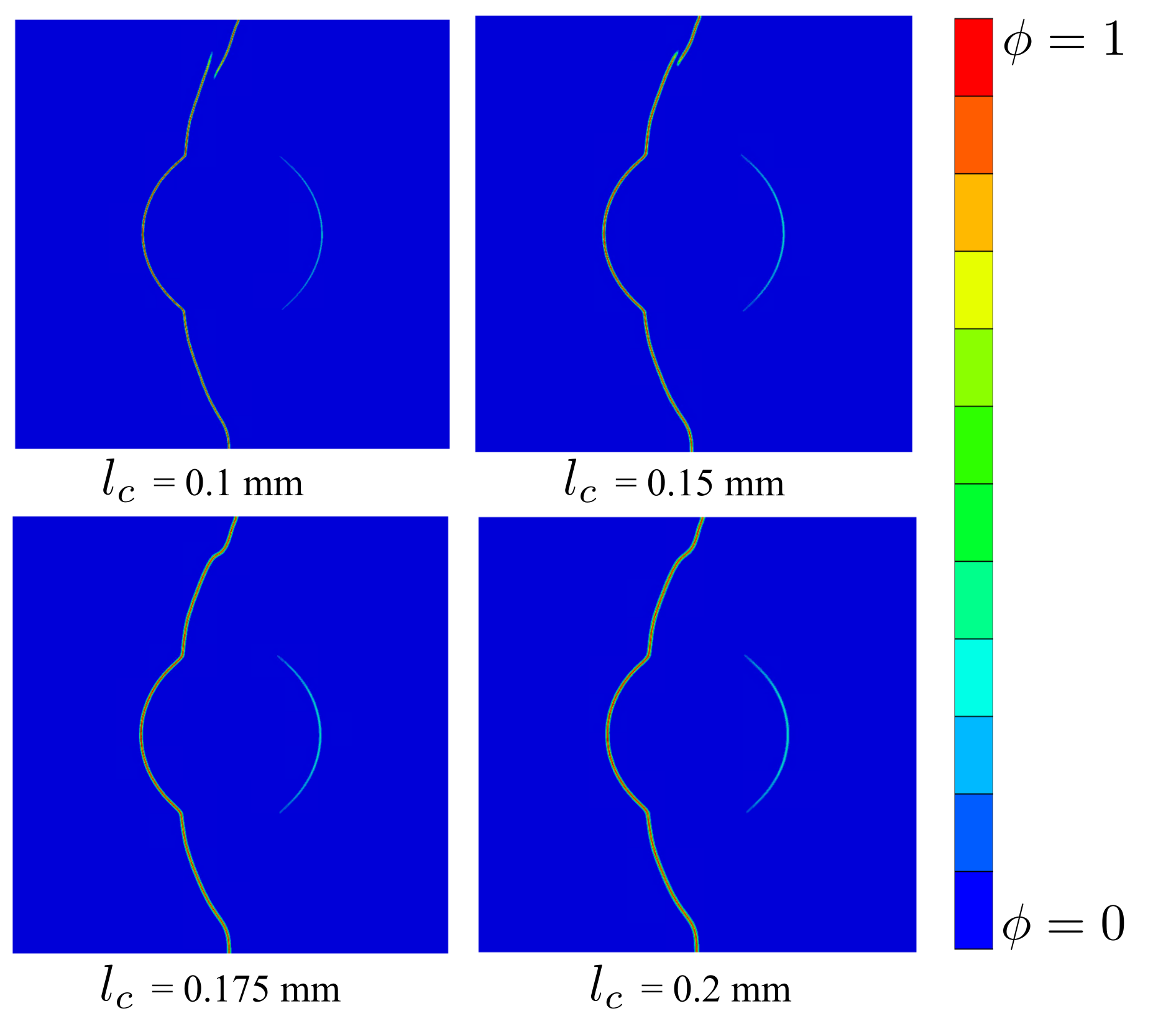}
    \caption{Fracture paths in the 2D single inclusion model have been simulated under tensile test conditions. Here, \(\phi\) denotes the phase-field fracture parameter, with \(\phi = 0\) indicating no damage, and \(\phi = 1\) illustrating a complete fracture.}
    \label{cpfm}
\end{figure}

\begin{figure}[t]
\center
    \includegraphics[width=0.6\textwidth]{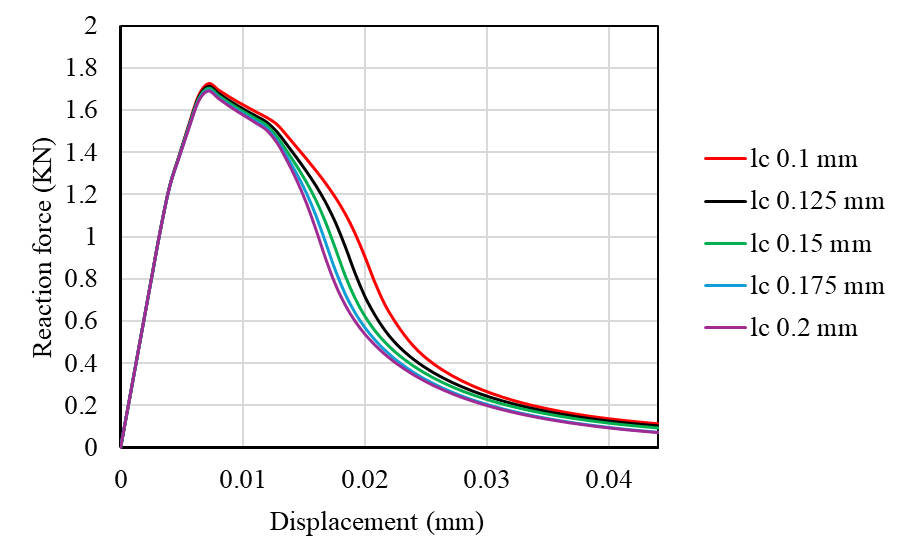}
    \caption{Sensitivity analysis of the parameter \(l_{c}\) conducted using the cohesive phase-field fracture model.}
    \label{lc_cpfm}
\end{figure}

In this study, a sensitivity analysis of the length scale parameter \(l_{c}\) is performed using a CPFM approach on the 2D benchmark model. All parameters are held constant except \(l_{c}\), which is systematically varied between 0.1 mm and 0.2 mm to assess its effect on the fracture paths and the reaction force-displacement relationship.

Figure \ref{cpfm} shows different fracture paths corresponding to different \(l_{c}\) values. As indicated by \cite{wu2017unified} and reflected in Figure~\ref{lc_cpfm}, two key observations emerge. First, the CPFM approach shows minimal sensitivity to \(l_{c}\), with changes in the parameter having a negligible effect on the reaction force-displacement response. This suggests that the proposed modifications have effectively converted \(l_{c}\) from a material parameter to a numerical one. Secondly, the reaction force-displacement curves exhibit cohesive behaviour, with a gradual decrease in reaction force after the softening zone, rather than a sudden decrease. This behaviour is particularly suitable for simulating semi-brittle materials such as concrete.

These results suggest that at smaller length scales, such as mesoscale or microscale, the \(l_{c}\) parameter no longer restricts the simulations. The CPFM approach therefore offers a more robust and flexible option for fracture simulations in multiphase materials at smaller scales.

%#########################################################################
\subsubsection{Hybrid model}
\begin{figure}[t]
\center
    \includegraphics[width=0.5\textwidth]{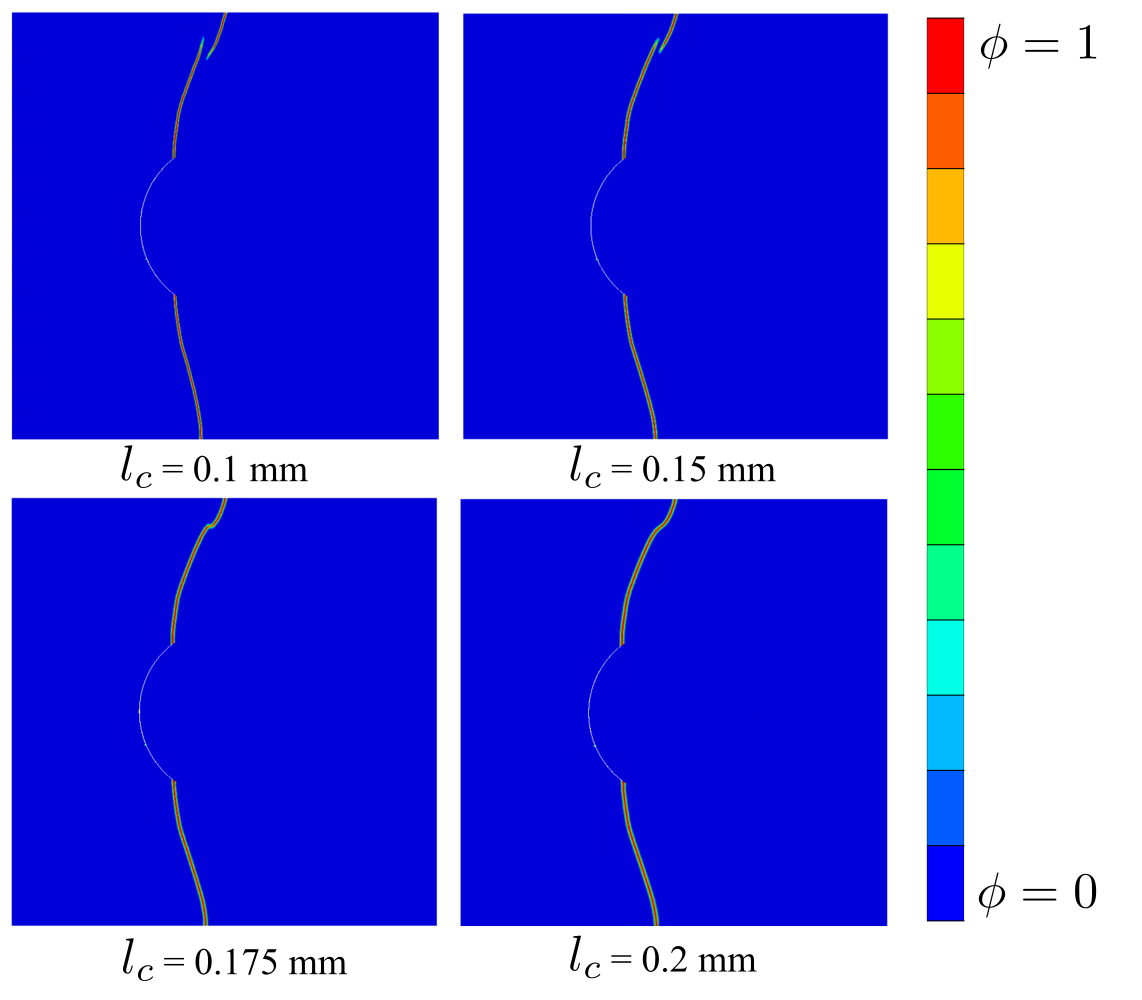}
    \caption{Fracture paths in the 2D single inclusion model have been simulated under tensile test conditions. CZM formulation is utilized in the interface zone and CPFM approach is utilized within mortar matrix and inclusion. Here, \(\phi\) denotes the phase-field fracture parameter, with \(\phi = 0\) indicating no damage, and \(\phi = 1\) illustrating a complete fracture.}
    \label{hybrid_fracture}
\end{figure}
\begin{figure}[t]
\center
    \includegraphics[width=0.6\textwidth]{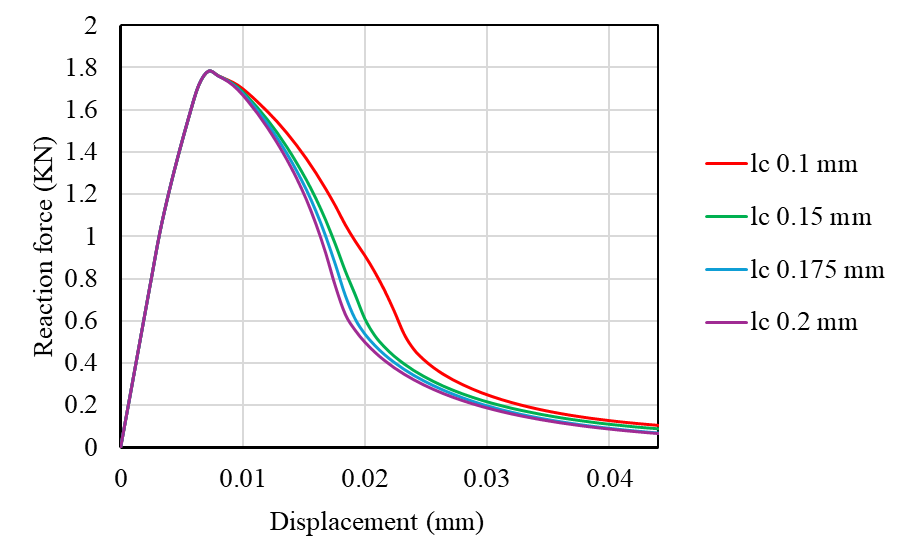}
    \caption{Sensitivity analysis of the parameter \(l_{c}\) conducted using the hybrid model, i.e. a combination of cohesive phase-field fracture and cohesive zone models.}
    \label{lc_hybrid}
\end{figure}
When the thickness of the interface zone in a composite material is much smaller than that of the other phases, the hybrid modelling technique becomes the most effective approach. This is particularly true for materials such as concrete at the meso or micro scale, where the interface is significantly thinner than the other phases.To accurately model such materials, it is essential to include the weak interface in numerical simulations. However, due to the small thickness of this zone in comparison to the larger surrounding phases, directly including it poses a major challenge. The interface zone requires extremely fine mesh elements for proper discretization, which significantly increases the number of elements and leads to a substantial rise in computational costs.

This challenge can be overcome by implementing CIEs within the interface zone. In the hybrid model of this study, the CPFM approach is applied to both the matrix and the inclusion, while the interface zone is represented by generating CIEs at the boundary between the inclusion and the matrix. In addition, a sensitivity analysis of \(l_c\) is performed using the hybrid formulation on a 2D benchmark model, with the results shown in Figure~\ref{hybrid_fracture} and Figure~\ref{lc_hybrid}. Similar to the CPFM method, the hybrid model shows insensitivity to \(l_c\) and cohesive behaviour in the reaction force-displacement curves.
%#########################################################################
\subsection{Evaluating Numerical Efficiency}
\begin{figure}[t]
\center
    \includegraphics[width=0.7\textwidth]{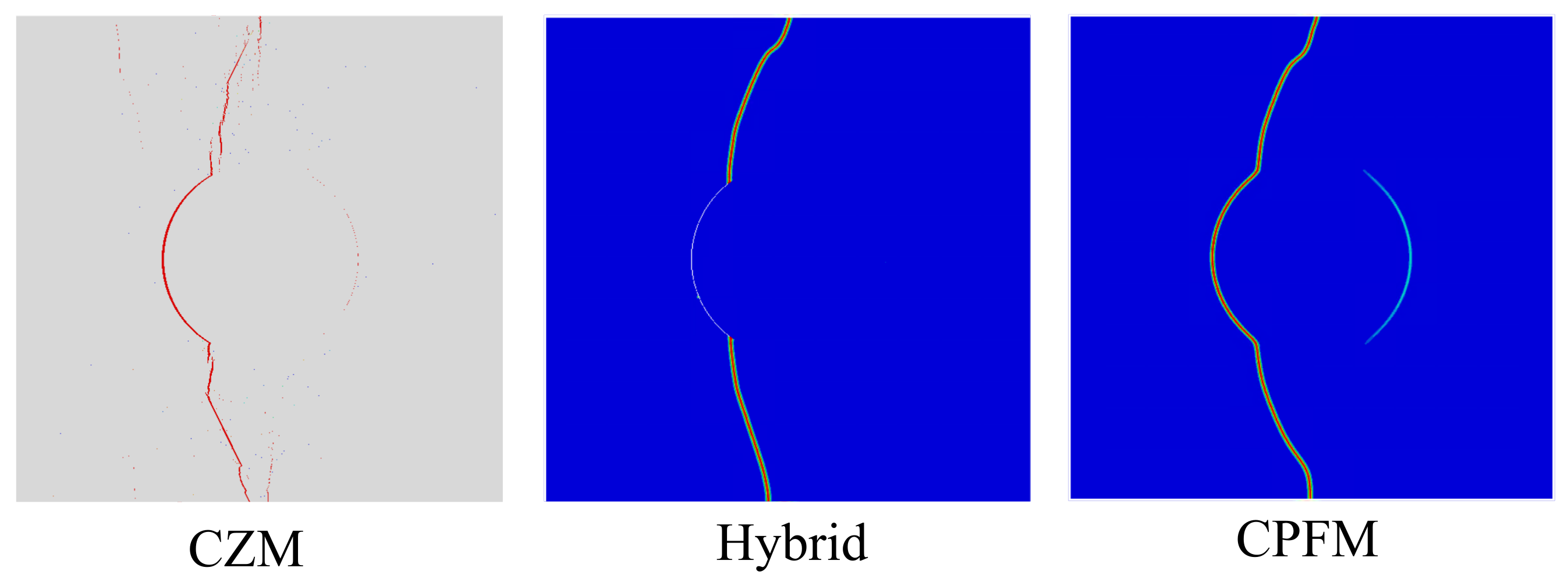}
    \caption{Simulated fracture paths by implementing the Cohesive Zone Model (CZM), Cohesive Phase-Field Fracture (CPFM), and a hybrid approach within a two-dimensional single inclusion model, under the boundary conditions illustrated in Figure~\ref{bc_gem}.}
    \label{comp_frac_paths}
\end{figure} 
\begin{figure}[t]
\center
    \includegraphics[width=\textwidth]{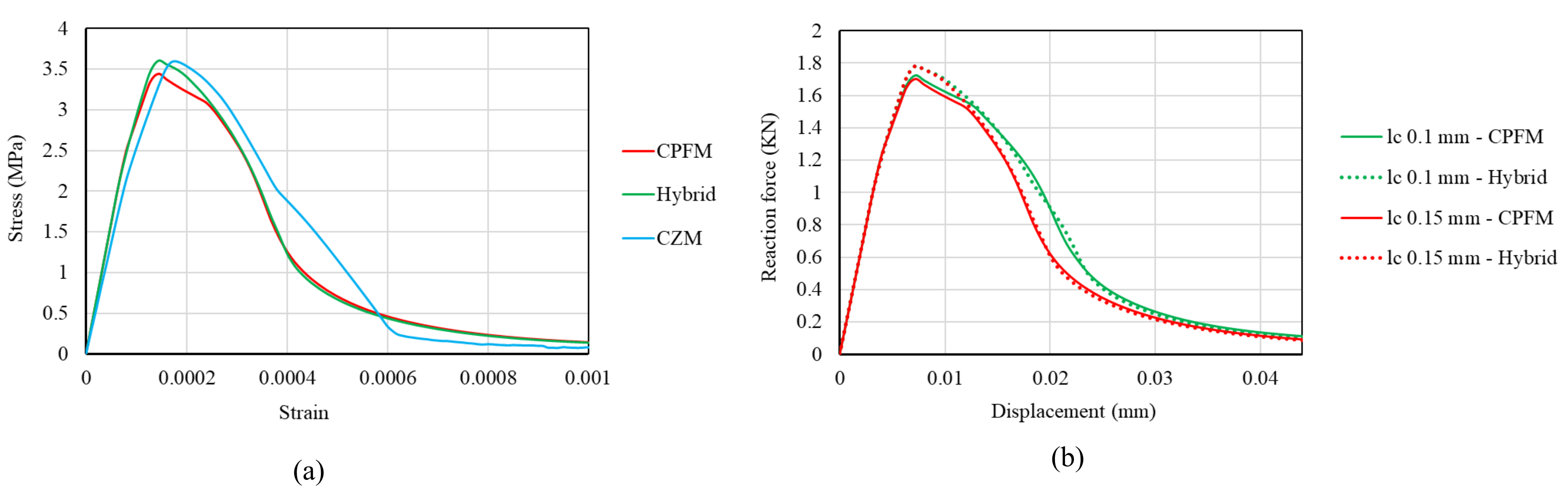}%
    \caption{A quantitative analysis is represented, featuring: (a) the reaction force-displacement curves acquired through finite element simulations of the two-dimensional single inclusion model, employing the CZM, CPFM, and the hybrid Approach; and (b) a comparative study of the \(l_{c}\) sensitivity analysis, conducted utilizing both the hybrid and the CPFM fracture models.}
    \label{comp_force}
\end{figure}
\begin{figure}[t]
\center
    \includegraphics[width=0.6\textwidth]{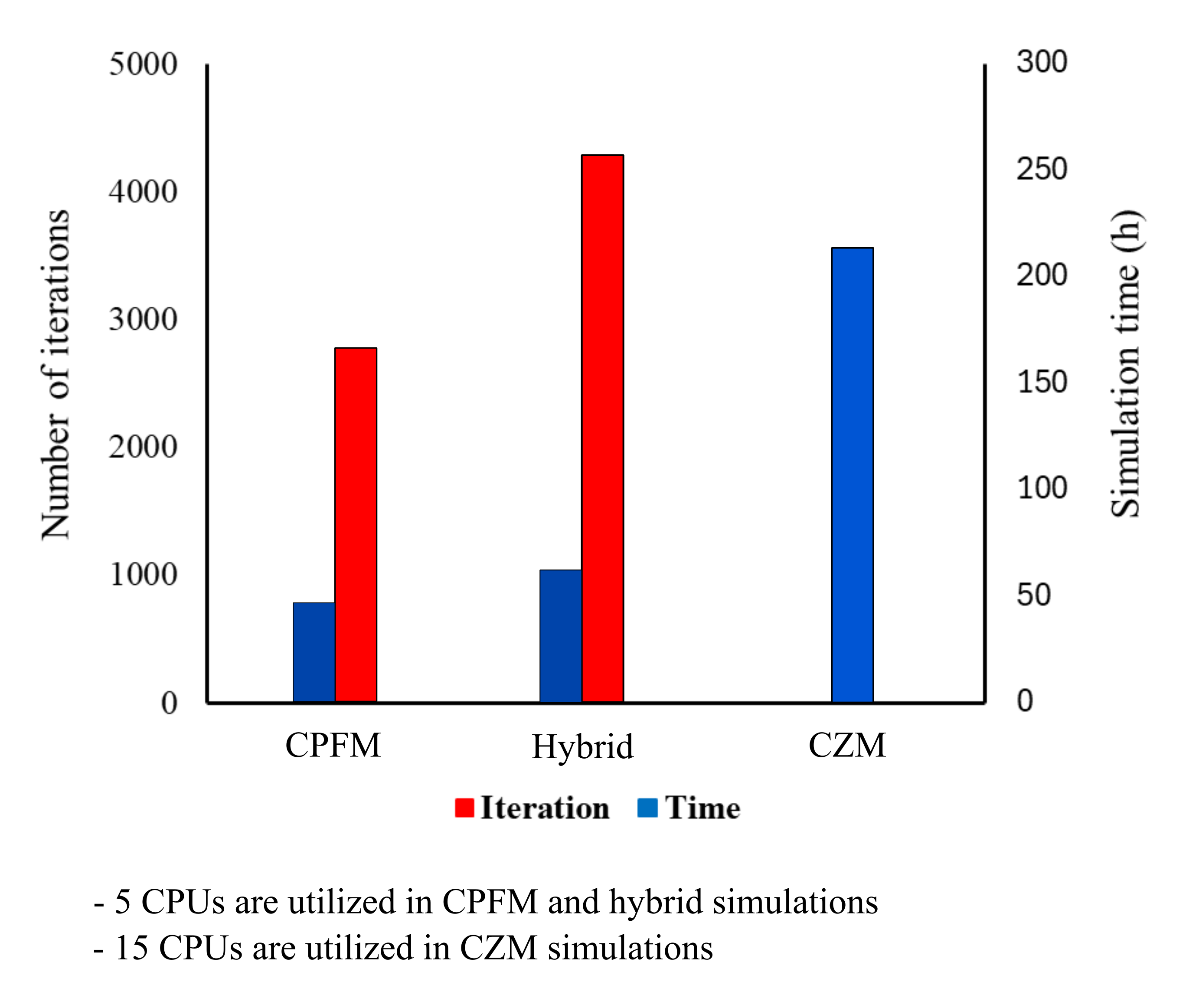}
    \caption{A comparative study focused on the numerical efficiency between the CPFM, CZM and the hybrid model.}
    \label{comp_iter}
\end{figure}
This section compares the presented methods for simulating fracture in multiphase materials at small length scales in terms of computational time and the total number of iterations required for a complete and successful finite element simulation. Due to the dependence of the length scale parameter on the material properties in the SPFM fracture model, and its limitations in small-scale simulations, this approach is not included in the comparisons. Figure~\ref{comp_frac_paths} illustrates the fracture paths obtained using the CZM, hybrid and CPFM approaches. Under similar boundary conditions applied to the benchmark model, all three methods produce comparable fracture paths. However, in terms of computational cost and number of iterations, there are some important differences between the models. These differences are discussed in more detail below.

\subsubsection{Cohesive zone model}
In the CZM simulation, the non-linear equation systems were solved using the Abaqus explicit solver. A mesh size of 0.05 mm was used, resulting in 180,151 solid elements and 361,390 CIEs for the 2D benchmark model. 

Figure \ref{comp_force}a presents the reaction force-displacement curve, comparing the CZM model with the hybrid and CPFM models. The results show a strong agreement between the models, with only a minor deviation observed in the softening zone. This behaviour is probably due to the need for careful calibration of numerous numerical parameters, in particular the TSL properties assigned to the CIEs in different phases. However, the main limitation of the CZM model is its very high computational cost. As can be observed in Figure~\ref{comp_iter}, by utilizing 15 CPUs, it took 258 hours and more that 33,000 increments to achieve a successful CZM simulation of the benchmark model. In contrast, the CPFM and hybrid simulations used only 5 CPUs with a much lower computational cost. Within the utilized workstation used for FE simulation, each CPU has \(2.43 \times 10^8\) FLOPS (Floating-Point Operations) per core per GHz.
\subsubsection{Hybrid model}
While the hybrid model is effective, it has certain limitations. One of the main challenges is its complexity. The model integrates two different fracture approaches, CZM and CPFM approach, within a single finite element framework. This requires additional pre-processing effort as it involves the development of two sets of tools. The first set is used to generate CIEs, which in this study is done using Python scripts. The second set involves the creation of a user-defined element subroutine that implements the cohesive phase-field fracture method within Abaqus/CAE.

In terms of computational cost, the hybrid model combines CZM and CPFM, which increases the number of iterations required for the numerical solver to converge at each time step (see Figure~\ref{comp_iter}). As a result, the integration of these two methods results in higher computational cost as more iterations are required for the solver to achieve convergence. 

\subsubsection{Cohesive phase-field fracture model}
As mentioned in previous section, one of the main advantages of CPFM approach is its insensitivity to the length scale parameter, which makes it physically appropriate for small-scale fracture simulations. As shown in Figure~\ref{comp_force}a, the reaction force-displacement curve from the CPFM approach closely matches that of the hybrid model. Further analysis in Figure~\ref{comp_force}b, comparing the reaction force-displacement curves for different values of \(l_{c}\), shows that the CPFM model consistently converges with the hybrid model. This convergence is beneficial as it allows the simpler and less computationally intensive CPFM model to be used in small-scale simulations without reducing accuracy. However, it is important to note that the observed convergence depends on the thickness of the interface and its material properties. In cases beyond the reference model in this study, a convergence study may be required to adjust the CPFM parameters including the interface thickness accordingly.
\begin{figure}[t]
  \centering
  \begin{subfigure}{0.25\textwidth}
   \includegraphics[width=\linewidth]{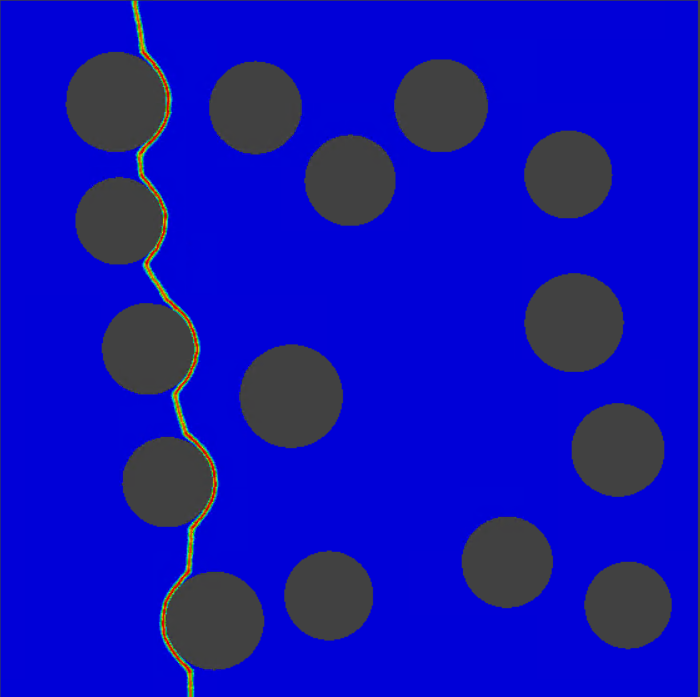} 
    \caption{Circular inclusions}
    \label{}
  \end{subfigure}
  \hspace{0.5em}  % Adjusted spacing
  % Fifth picture (Placeholder)
  \begin{subfigure}{0.25\textwidth}
    \includegraphics[width=\linewidth]{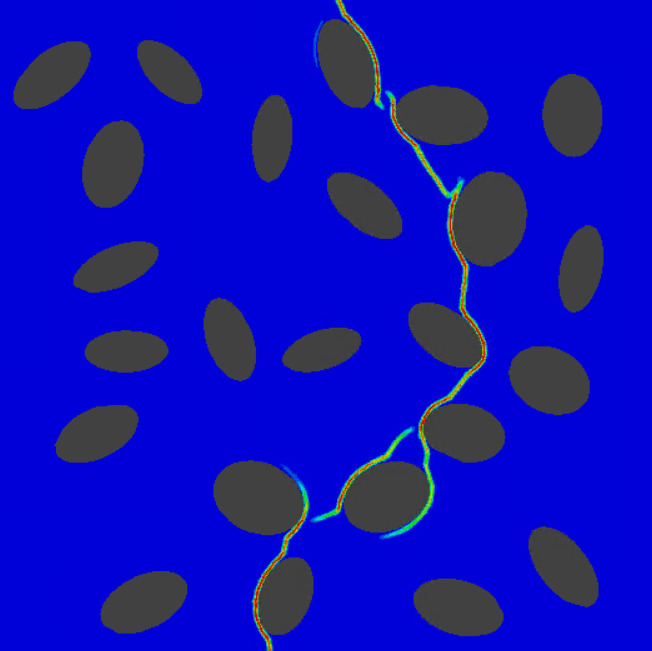} 
    \caption{Elliptical inclusions}
    \label{e}
  \end{subfigure}
  \hspace{0.5em}  % Adjusted spacing
  % Sixth picture (Placeholder)
  \begin{subfigure}{0.25\textwidth}
    \includegraphics[width=\linewidth]{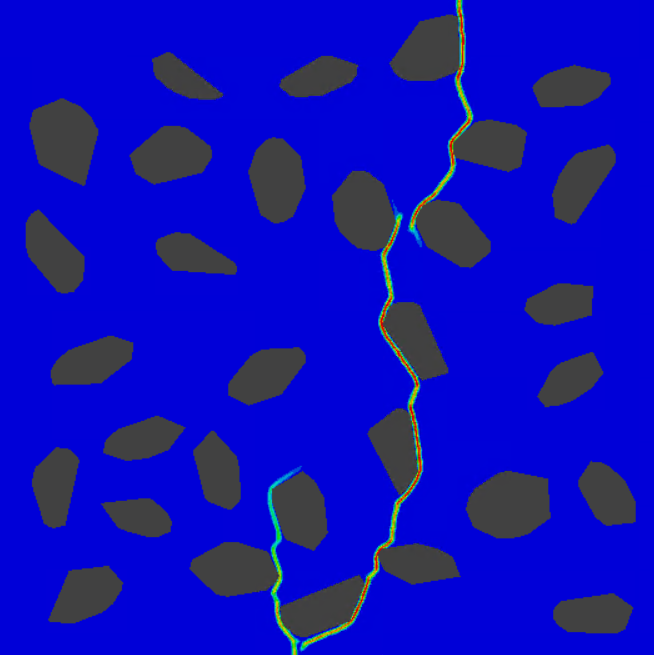} 
    \caption{Polyhedron inclusions}
    \label{f}
  \end{subfigure}  
  \caption{The study investigates fracture paths in complex models featuring multiple inclusions with interface regions. The model is 50 mm $\times$ 50 mm.  For these simulations, the CPFM technique is employed. The specific boundary conditions applied in the finite element analysis are illustrated in Figure~\ref{bc_gem}.}
  \label{meso_concrete}
\end{figure}
\begin{figure}[H]
\center
    \includegraphics[width=0.6\textwidth]{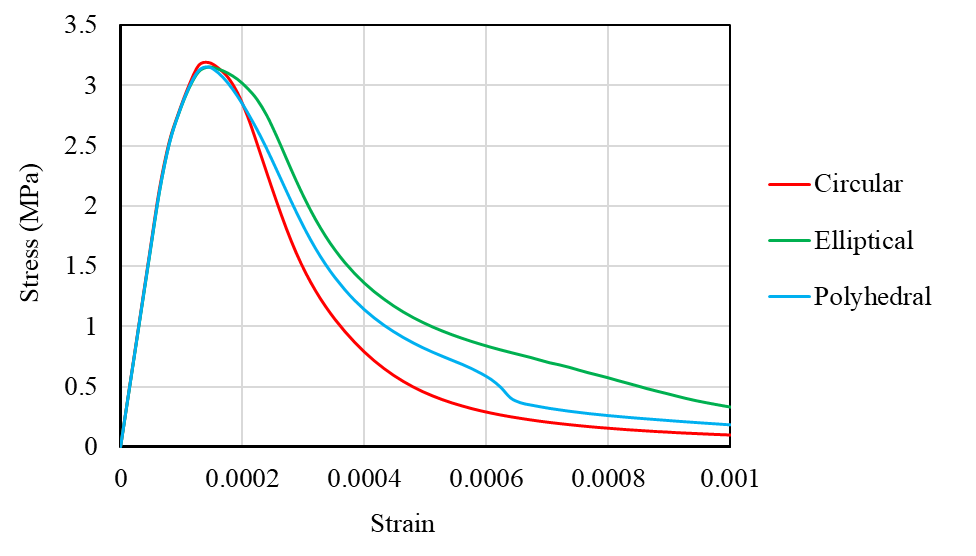}
    \caption{Reaction force-displacement curve from finite element simulations showcasing models with varied inclusion shapes and random spatial distribution of inclusions.}
    \label{concrete_graph}
\end{figure}
As can be observed in Figure \ref{comp_iter} the CPFM approach is more computationally efficient than other fracture models. Overall, the results indicate that the CPFM fracture method is the preferred technique for simulating fracture in multiphase materials, especially at smaller scales, provided that the thickness of the interface is relatively large compared to the other phases. In scenarios where the interface thickness is very small, such as the ITZ in concrete materials, the thickness can be artificially increased and the material properties scaled to make CPFM a fracture approach that is insensitive to interface width, as discussed in \cite{zhou2022interface}.

%#########################################################################
\subsection{Dimensional progression}
In this section, the expansion of the CPFM approach to encompass more complex conditions is explored through the presentation of various finite element simulations. 
\subsubsection{Complex 2D Simulations}
The simulated fracture paths in Figure~\ref{meso_concrete} show 2D models with a stochastic distribution of inclusions accounting for 20\% of the total volume, each with a 0.5 mm interface layer thickness. Finite element simulations successfully reproduce complete fracture paths in models with multiple inclusions. Figure~\ref{concrete_graph} provides a comparative analysis of the reaction force-displacement curves from each model.

\subsubsection{3D simulations}
\begin{figure}[H]
  \centering
  \begin{subfigure}{0.2\textwidth}
    \includegraphics[width=\linewidth]{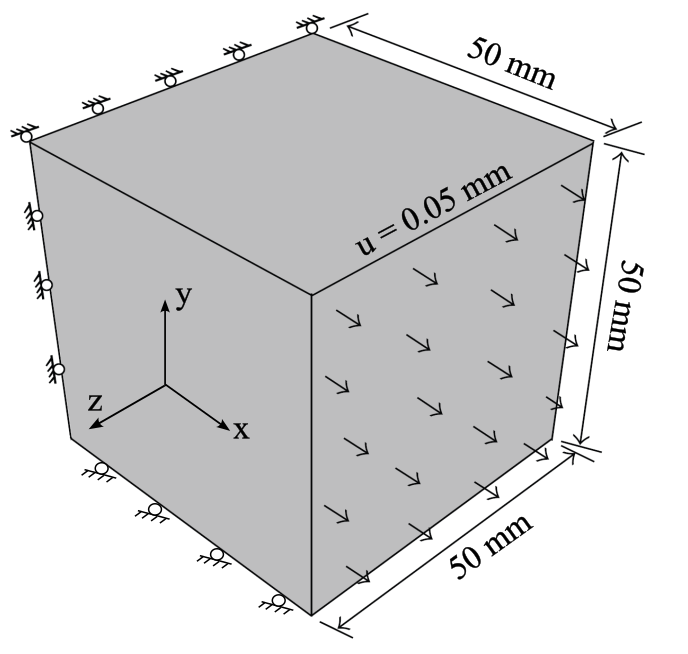}
    \caption{}
    \label{3D_dims}
  \end{subfigure}
  \hspace{1em}  % Adjust this value to change the spacing. Set to 0em for no space.
  \begin{subfigure}{0.2\textwidth}
    \includegraphics[width=\linewidth]{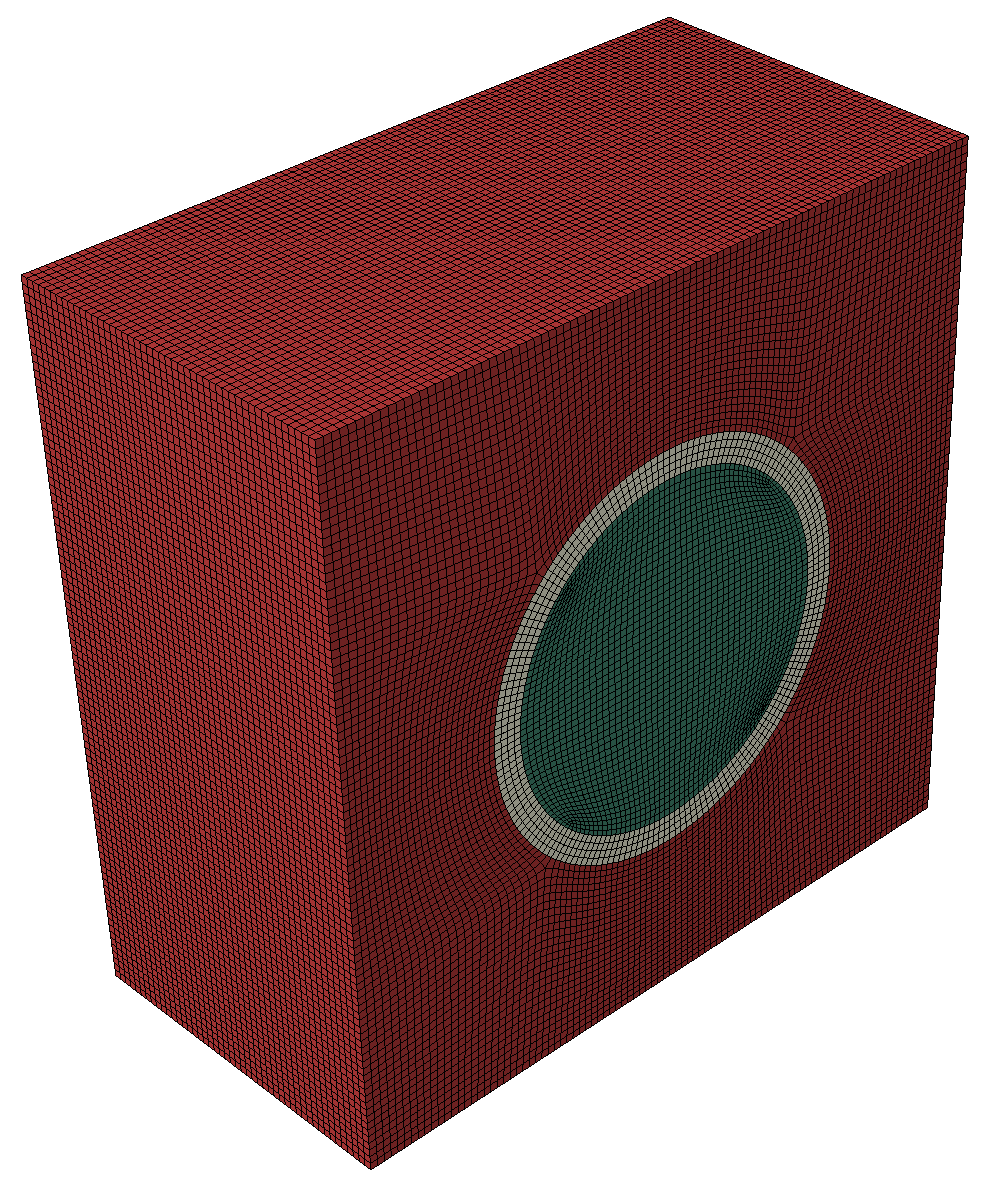}
    \caption{}
    \label{figure_3b}
  \end{subfigure}
  \caption{(a) Specified dimensions and boundary conditions are on the 3D specimen. (b) The 3D specimen incorporates a spherical inclusion surrounded by an interface region. To depict the phases, the model is halved in the x direction. }
  \label{2D}
\end{figure}

In this section, the capability of the CPFM fracture approach in simulating 3D boundary value problems for multiphase materials is demonstrated. A 3D model with a single inclusion, which has boundary conditions analogous to the 2D model, is chosen to represent the initiation and propagation of cracks. The dimensions and imposed boundary conditions are depicted in Figure~\ref{3D_dims}. In Figure~\ref{figure_3b}, a cross-section of half the model along the x-axis is shown to illustrate the three distinct phases. 

\begin{figure}[H]
  \centering
  \begin{subfigure}{0.2\textwidth}
    \includegraphics[width=\linewidth]{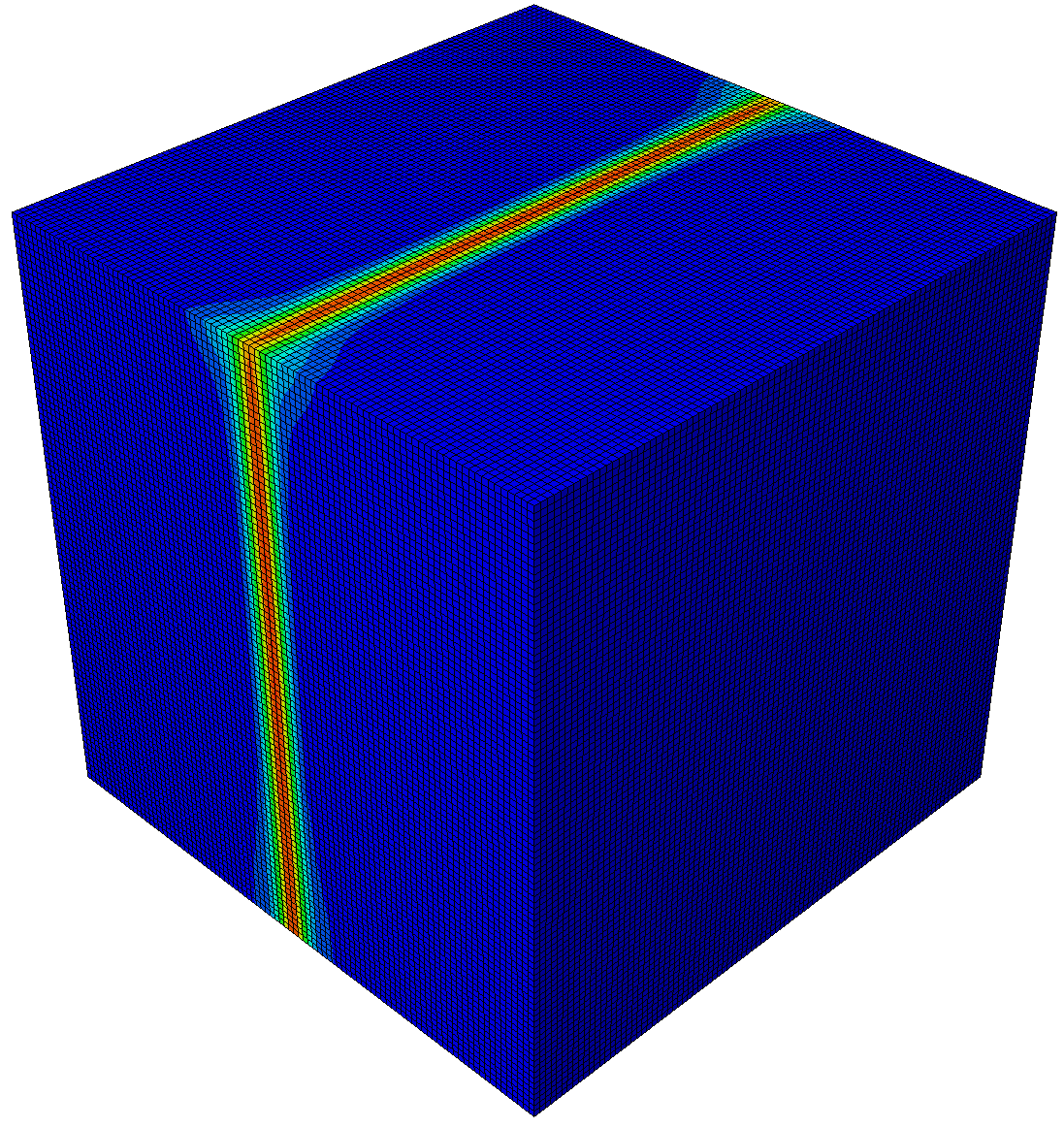}
    \caption{}
    \label{}
  \end{subfigure}
  \hspace{1em}  % Adjust this value to change the spacing. Set to 0em for no space.
  \begin{subfigure}{0.2\textwidth}
    \includegraphics[width=\linewidth]{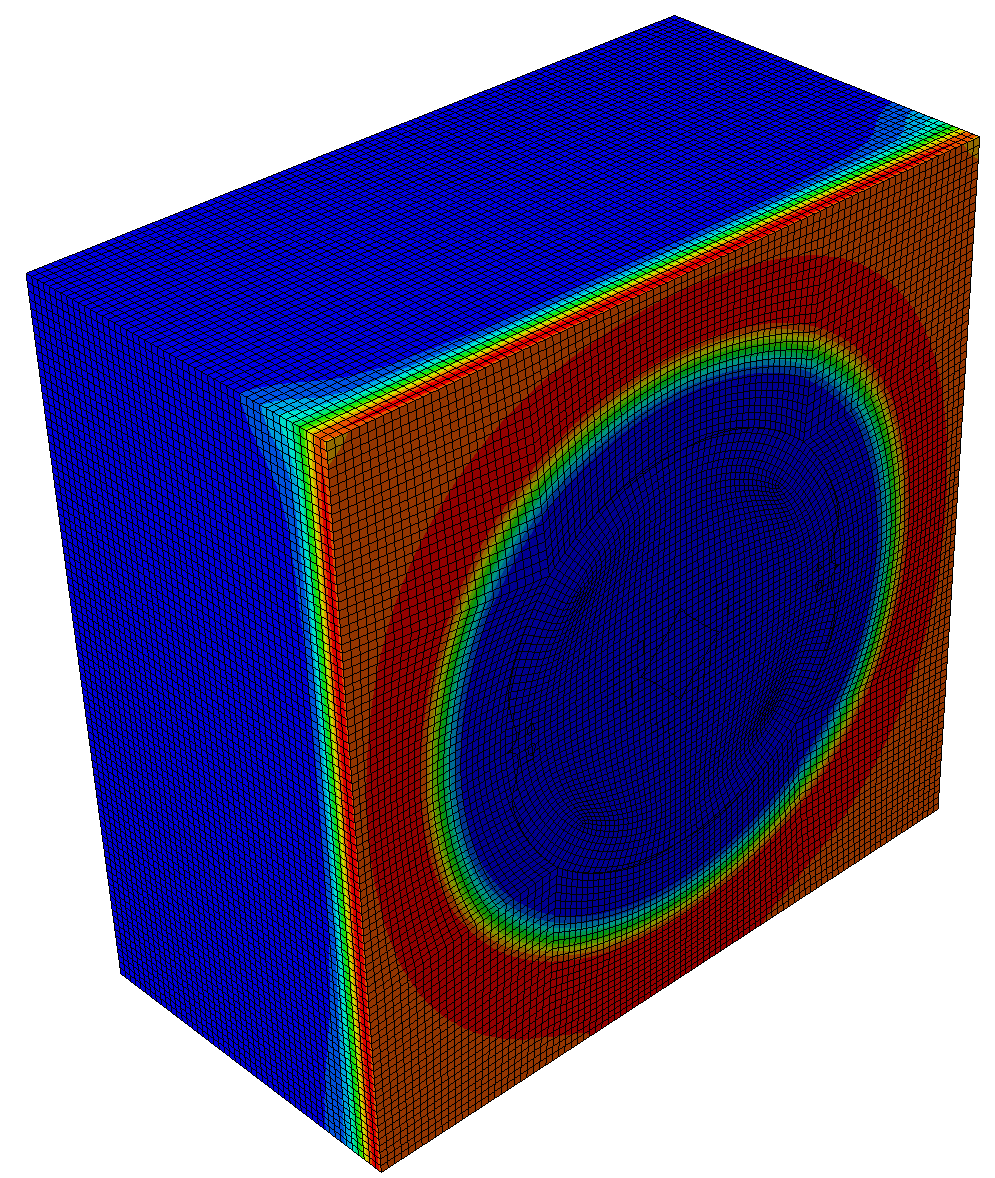}
    \caption{}
    \label{}
  \end{subfigure}
  \caption{Fracture of a single 3D inclusion under the boundary condition shown in Figure~\ref{3D_dims}. The images are extracted at the end of the simulation time.}
  \label{3D}
\end{figure} 

\begin{figure}[H]
\center
    \includegraphics[width=0.8\textwidth]{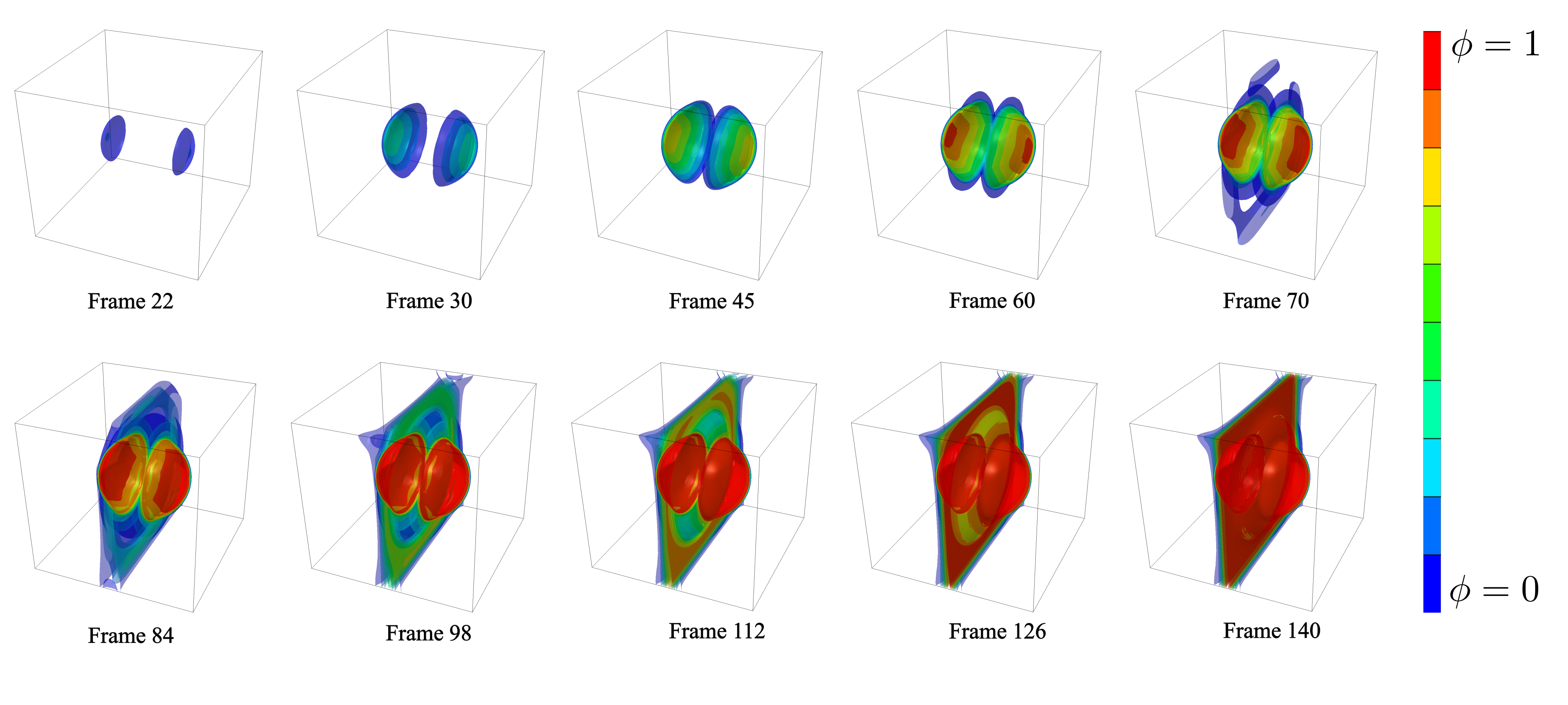}
    \caption{Initiation and propagation of fracture are modeled within a 3D single inclusion model. The cohesive phase-field fracture approach is utilized for the numerical simulation, which comprises a total of 140 frames.}
    \label{iso}
\end{figure}
\begin{figure}[H]
\center
    \includegraphics[width=0.6\textwidth]{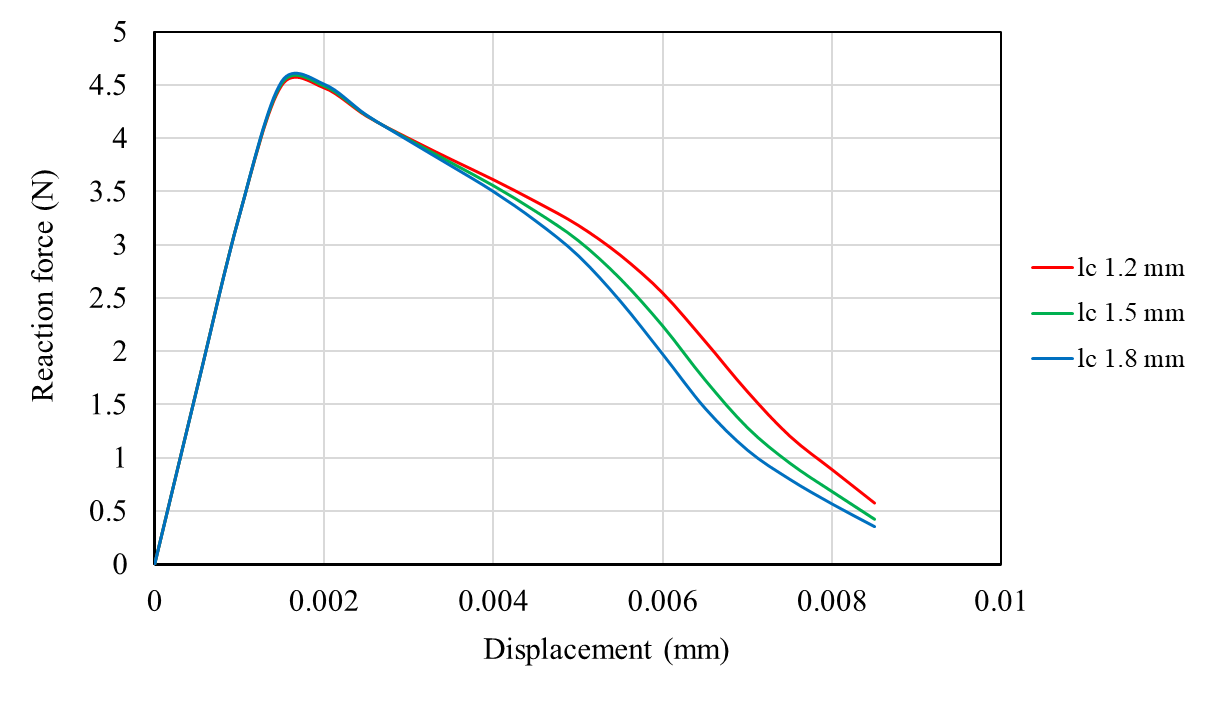}
    \caption{\(l_{c}\) sensitivity analysis conducted on 3D single inclusion specimen using the cohesive phase-field fracture approach.}
    \label{3D_lc_sens}
\end{figure}
Figure \ref{3D} illustrates the simulation outcomes using the CPFM approach for the 3D problem. Areas colored in blue indicate no damage (\(\phi=0\)), whereas the red regions signify damaged areas with \(\phi > 0.95\). The isosurface representation of the damage phase-field, showcasing the initiation and progression of the crack, is provided in Figure~\ref{iso}. As discerned from Figure~\ref{iso}, damage initiation occurs within the interface zone and subsequently propagates through the matrix.

Figure \ref{3D_lc_sens} clearly demonstrates the CPFM's insensitivity to variations in \(l_{c}\) in 3D simulations. Notably, changes in \(l_{c}\) result in slight changes in force-displacement curves.

%#########################################################################
\subsection{Microstructural Optimization Pathway}
This section provides a comprehensive analysis of how interfacial mechanical properties, geometric features of inclusions and voids affect the fracture properties of particulate composites. The study shows that the spatial distribution of inclusions influences the fracture properties of multiphase materials. The main purpose of this section is to highlight the importance of the spatial distribution of inclusions, interface properties and the presence of voids within the microstructure in optimising composite design, which are often overlooked in many studies. 
\subsubsection{Impact of Inclusion Spatial Distribution}  
\begin{figure}[H]
  \centering
  % First column
  \begin{subfigure}{0.222\textwidth}
    \includegraphics[width=\linewidth]{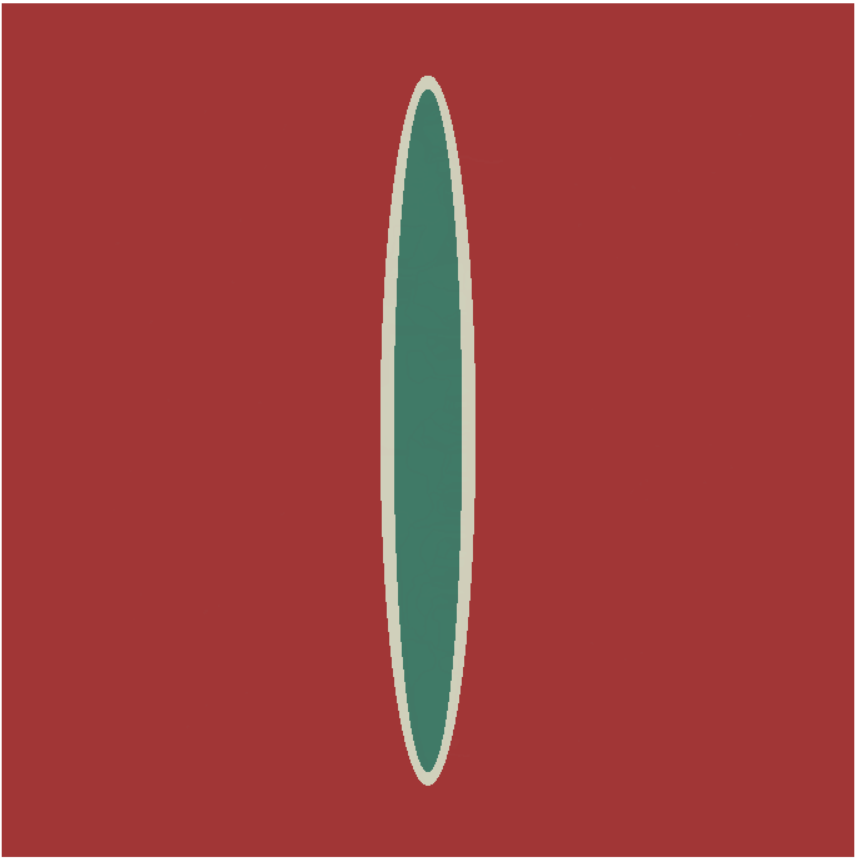} \\
        \vspace{0.05em} % distance between the two pictures
    \includegraphics[width=\linewidth]{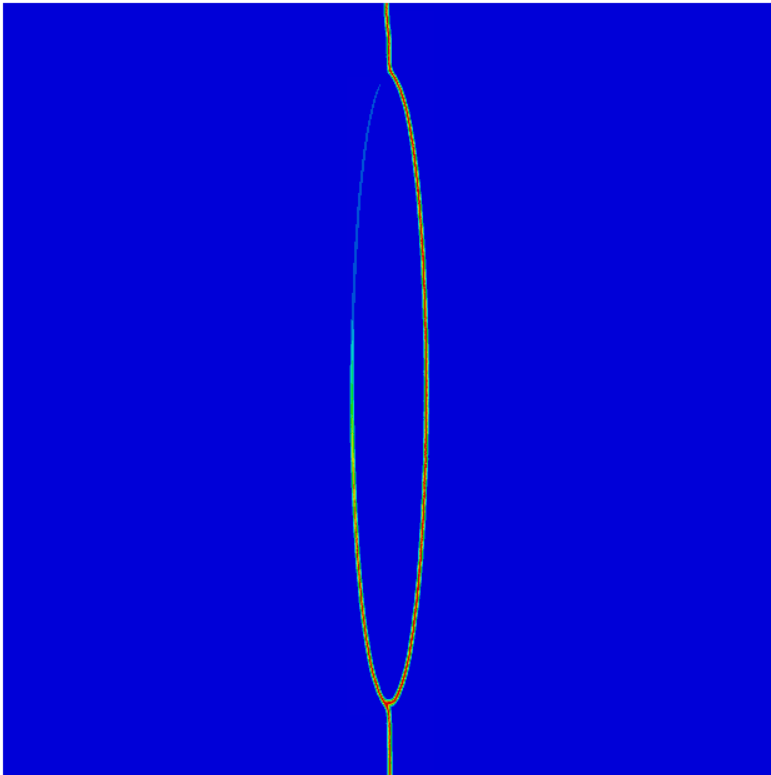} 
    \caption{}
    \label{type1}
  \end{subfigure}
  \hspace{0.5em}
  % Second column
  \begin{subfigure}{0.222\textwidth}
    \includegraphics[width=\linewidth]{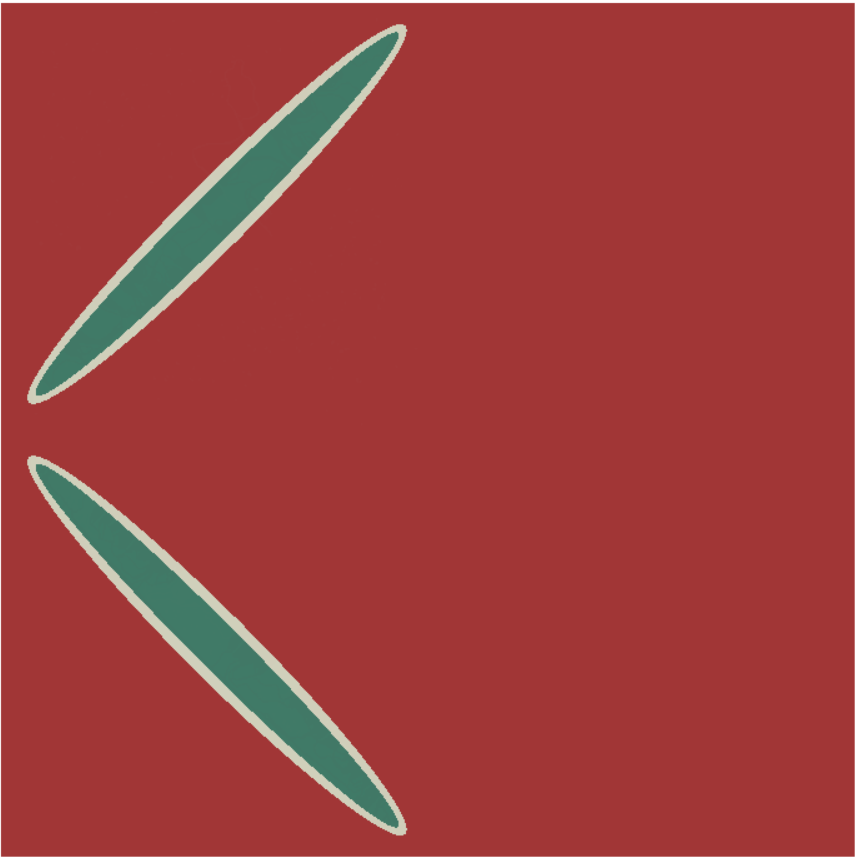} \\
        \vspace{0.05em} % distance between the two pictures
    \includegraphics[width=\linewidth]{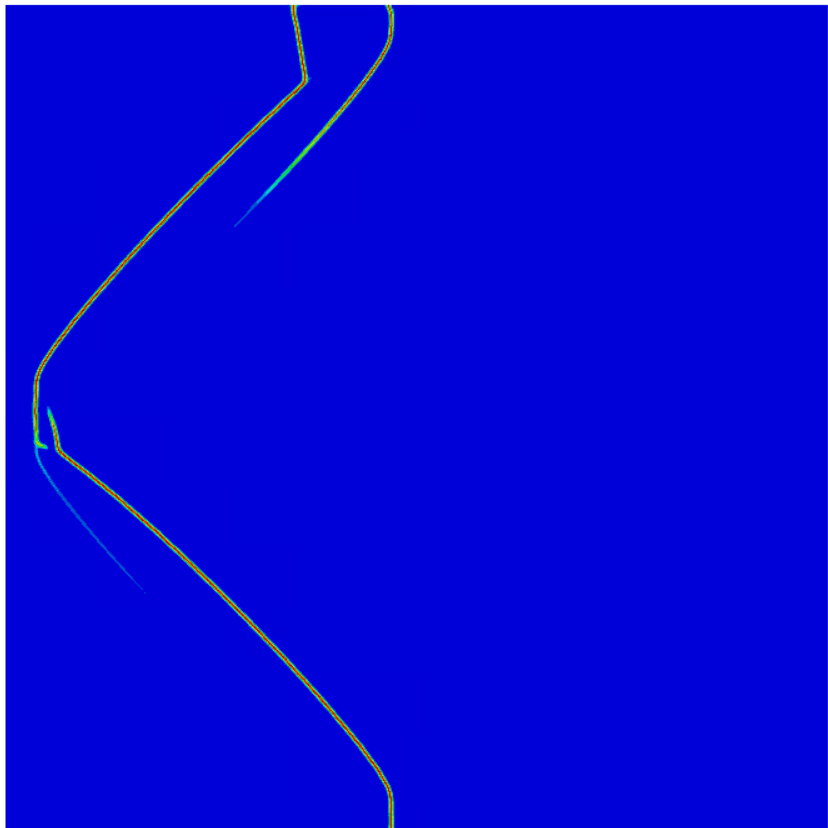} 
    \caption{}
    \label{type2}
  \end{subfigure}
 \hspace{0.5em}
  % Third column
  \begin{subfigure}{0.222\textwidth}
    \includegraphics[width=\linewidth]{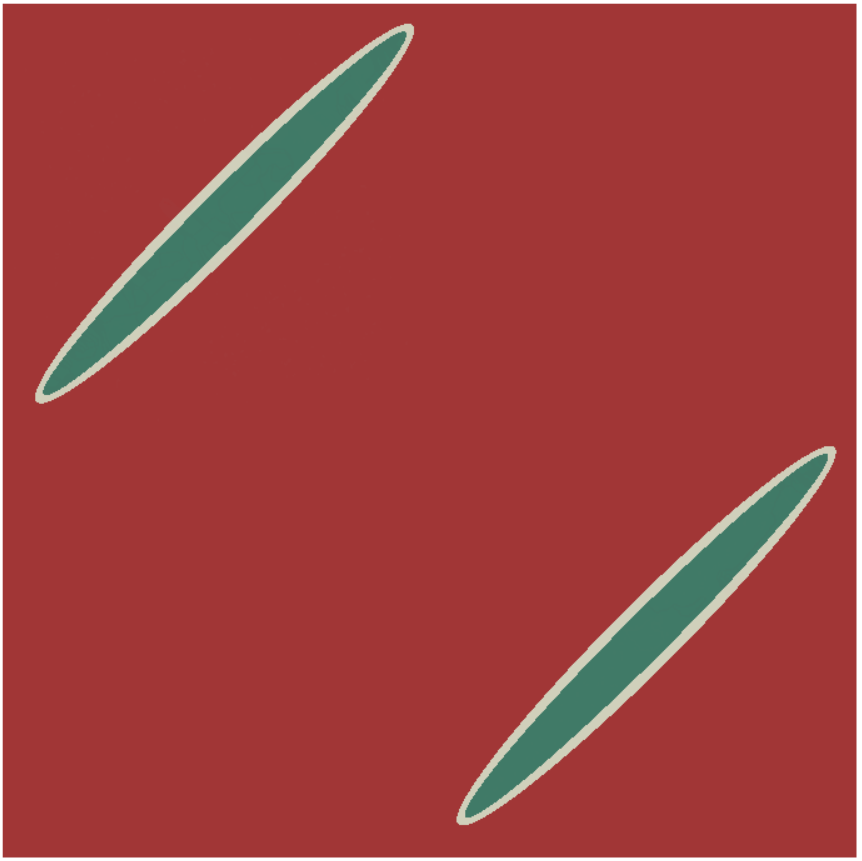} \\
        \vspace{0.05em} % distance between the two pictures
    \includegraphics[width=\linewidth]{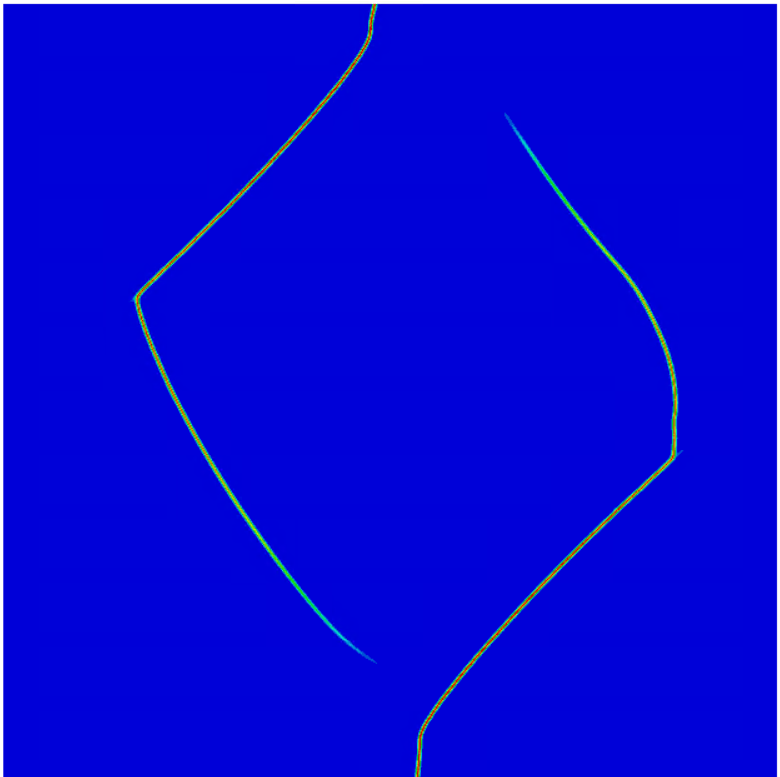} 
    \caption{}
    \label{type3}
  \end{subfigure}
  \hspace{0.5em}
  % Fourth column
  \begin{subfigure}{0.222\textwidth}
   \includegraphics[width=\linewidth]{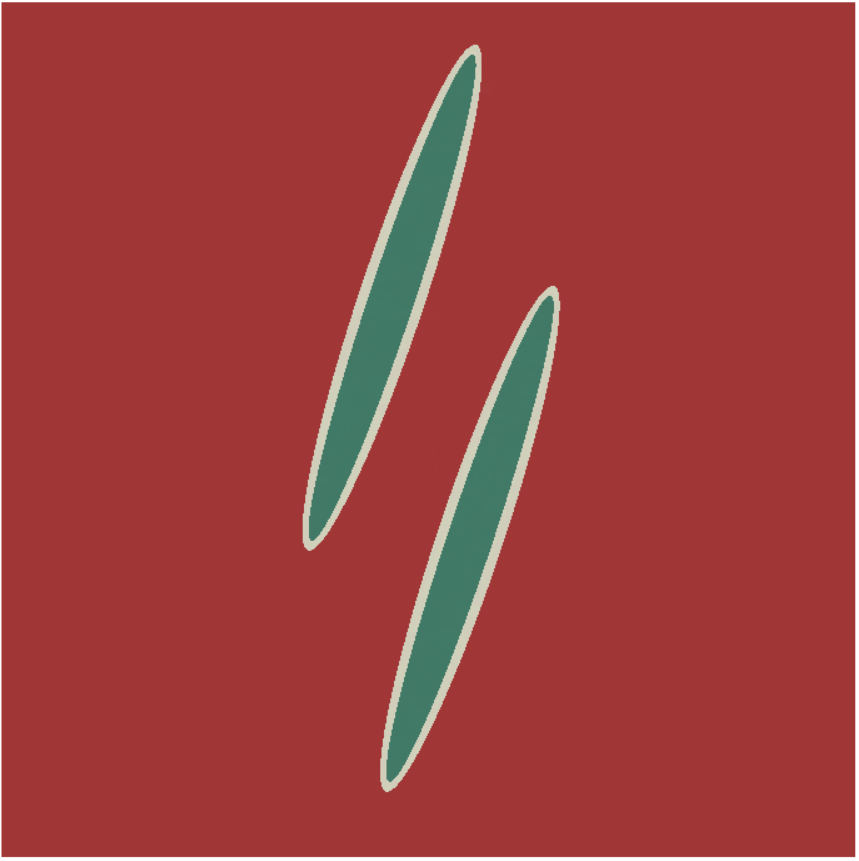} \\
    \vspace{0.05em} % distance between the two pictures
    \includegraphics[width=\linewidth]{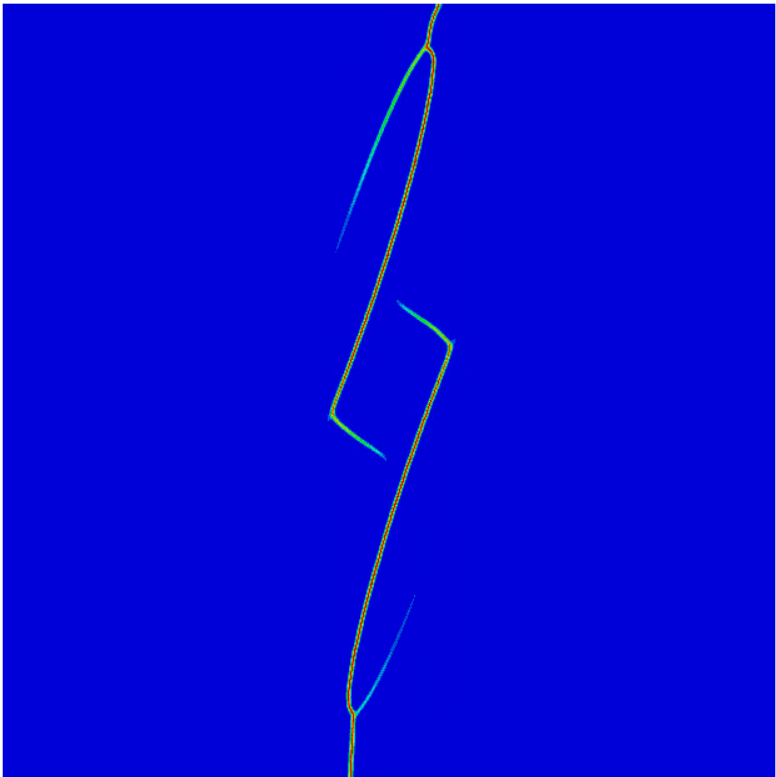} 
    \caption{}
    \label{type4}
  \end{subfigure}
  \caption{Effect of number of inclusions and their spatial distribution on simulated fracture paths. Two elliptical inclusion samples possess the identical volume fraction of inclusion and interface phase, mirroring the volume fractions in the single inclusion model.}
  \label{Inc_topology_crack}
\end{figure}

The influence of inclusion distribution is investigated using four unique microstructures with ellipsoidal aggregate particles, shown in Figure~\ref{Inc_topology_crack}. Each model consists of three phases: matrix, inclusion and interface zone, with the same volume fraction for inclusion and interface, and is discretised with a uniform mesh size of 0.05 mm. The boundary conditions for the finite element analysis are shown in Figure~\ref{bc_gem}.

\begin{figure}[H]
\center
    \includegraphics[width=0.6\textwidth]{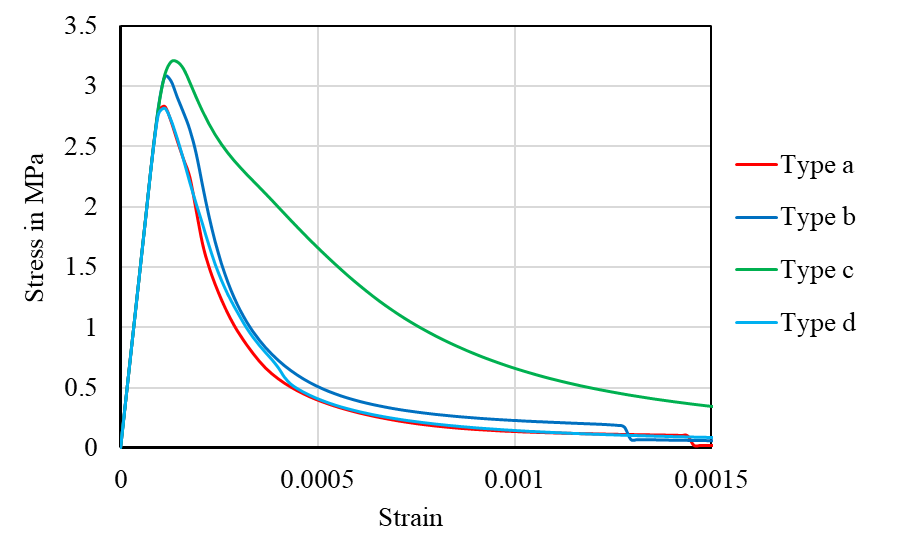}
    \caption{Calculated stress-strain curve derived from the models illustrated in Figure~\ref{Inc_topology_crack}}
    \label{topology_property}
\end{figure}

As shown in the second row of Figure~\ref{Inc_topology_crack}, the distribution of the inclusions, and consequently the interface, has a significant effect on the simulated fracture paths. This effect is particularly evident in Figure~\ref{type3}, where the strategic placement of two aggregate particles causes the entire fracture to occur through two primary cracks, improving the fracture toughness of the material. As can be seen in Figure~\ref{topology_property}, the dissipated fracture energy in Figure~\ref{type3} is approximately 45\% higher that of the single inclusion model in Figure~\ref{type1}, demonstrating the potential for improving fracture properties through strategic inclusion placement.

\begin{figure}[H]
  \centering
  \begin{subfigure}{0.222\textwidth}
    \includegraphics[width=\linewidth]{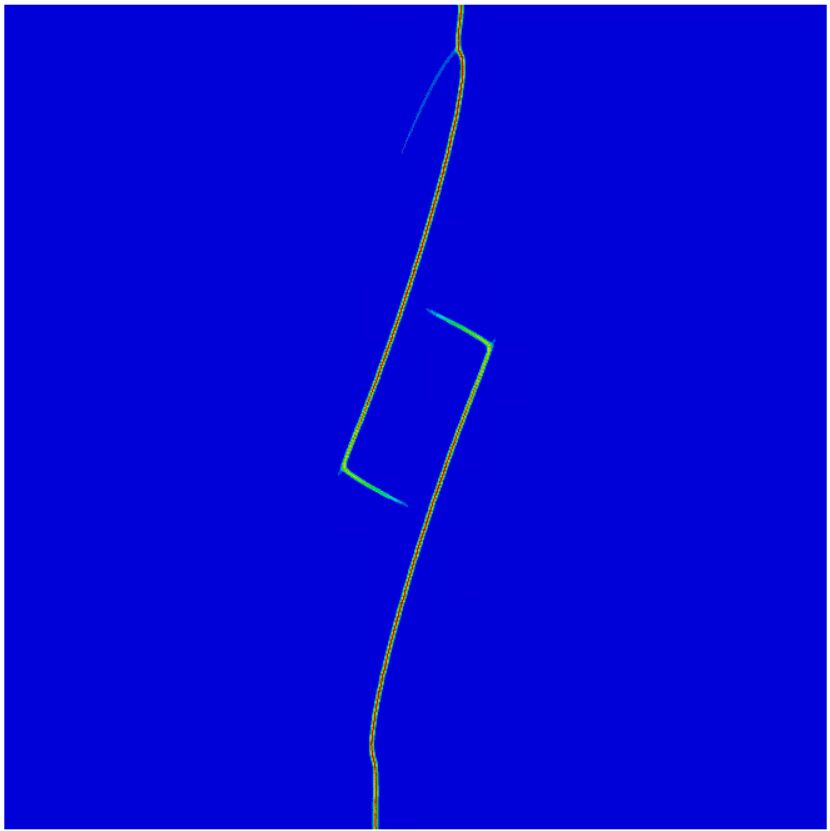} 
    \caption{Interface 1}
    \label{}
  \end{subfigure}
  \hspace{0.5em}  % Adjusted spacing
  % Fifth picture (Placeholder)
  \begin{subfigure}{0.222\textwidth}
    \includegraphics[width=\linewidth]{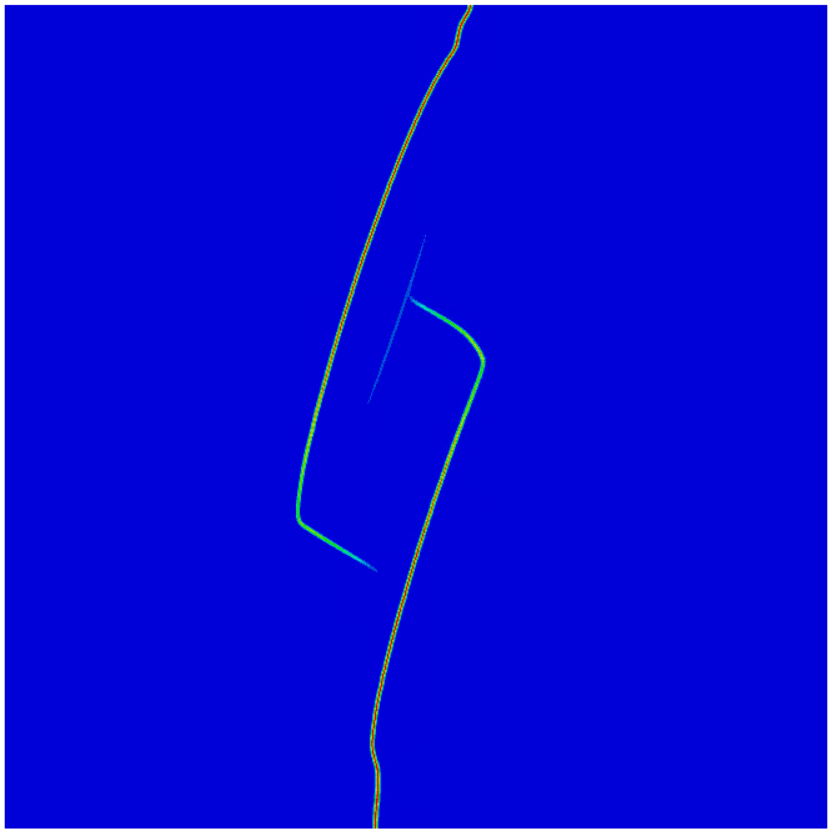} 
    \caption{Interface 2}
    \label{e}
  \end{subfigure}
  \hspace{0.5em}  % Adjusted spacing
  % Sixth picture (Placeholder)
  \begin{subfigure}{0.222\textwidth}
    \includegraphics[width=\linewidth]{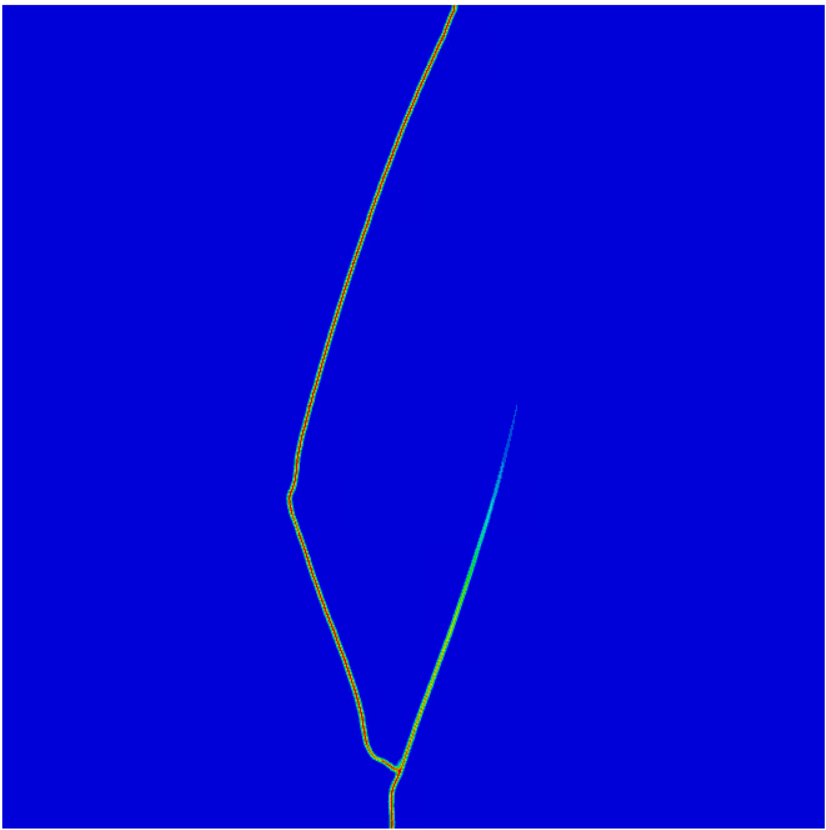} 
    \caption{Interface 3}
    \label{f}
  \end{subfigure}  
   \begin{subfigure}{0.222\textwidth}
    \includegraphics[width=\linewidth]{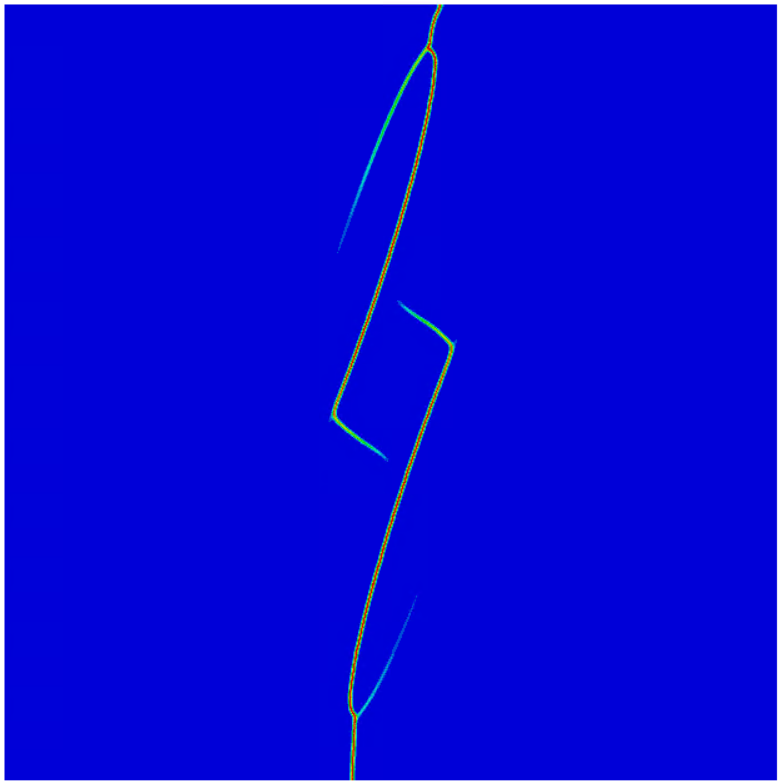} 
    \caption{Reference}
    \label{f}
  \end{subfigure}   
  \caption{Fracture paths in three unique microstructures that have an equivalent volume fraction of inclusion and interface phases. The material properties of the interface are provided in Table \ref{itz_prop}.}
  \label{meso_opt_frac}
\end{figure}

\subsubsection{Impact of Voids and Interface Material Properties}
\begin{figure}[t]
\center
    \includegraphics[width=0.6\textwidth]{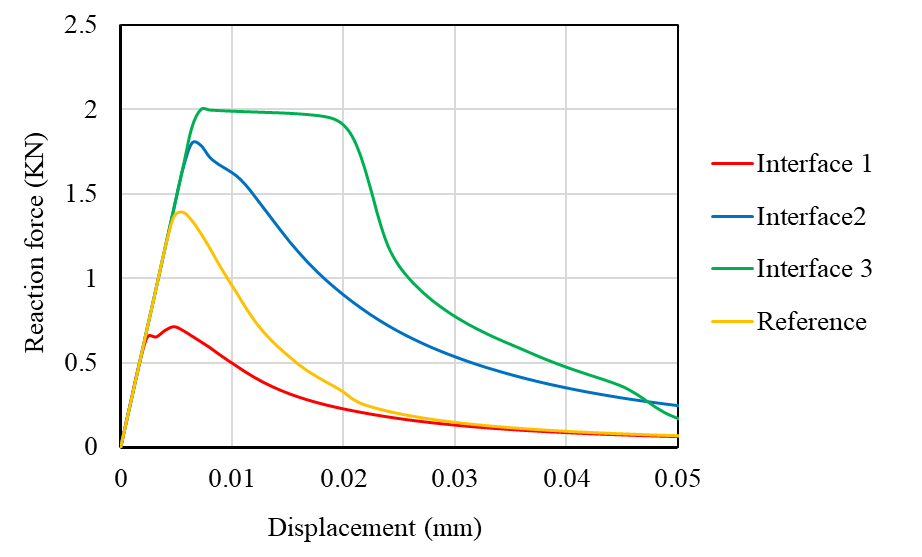}
    \caption{Force-displacement curves derived from FE simulations using distinct Interface properties as detailed in Table \ref{itz_prop}. While other material attributes are kept constant as specified in Table \ref{PFM_sim}, only the properties of the interface are varied.}
    \label{ITZ_graph}
\end{figure}

To evaluate the influence of interface properties on the fracture behaviour of composite samples, three different new fracture properties for the interface zone are detailed in Table \ref{itz_prop}. Interface 1 represents weakened properties while interface 3 represents strongest interface properties. Figure~\ref{meso_opt_frac} illustrates the fracture path for the two inclusion scenario and shows that changing the interface properties deflects the fracture path. Figure~\ref{ITZ_graph} shows that these variations increase the peak reaction force and energy dissipation during failure.

\begin{table}[t]
\caption{Material properties of interface zone. Interface 1 signifies extremely weak interface attributes and Interface 3 denotes highly robust interface characteristics.}\label{itz_prop}
	\begin{adjustwidth}{-\extralength}{0cm}
		\newcolumntype{C}{>{\centering\arraybackslash}X}
		\begin{tabularx}{\fulllength}{lCC}
			\toprule
			\textbf{Phase} & \makecell{\textbf{Fracture energy} \\ \boldmath$G_{c}$ \textbf{(N/mm)}} & \makecell{\textbf{Failure strength} \\ \boldmath$\sigma_{u}$ \textbf{(MPa)}} \\
			\midrule
			Reference     & 0.02 & 2.4 \\
            Interface 1   & 0.01 & 1 \\
			Interface 2   & 0.06  & 3.2 \\
			Interface 3   & 0.1  & 4 \\
			\bottomrule
		\end{tabularx}
	\end{adjustwidth}
\end{table}

\begin{figure}[H]
  \centering
  \begin{subfigure}{0.222\textwidth}
\includegraphics[width=\linewidth]{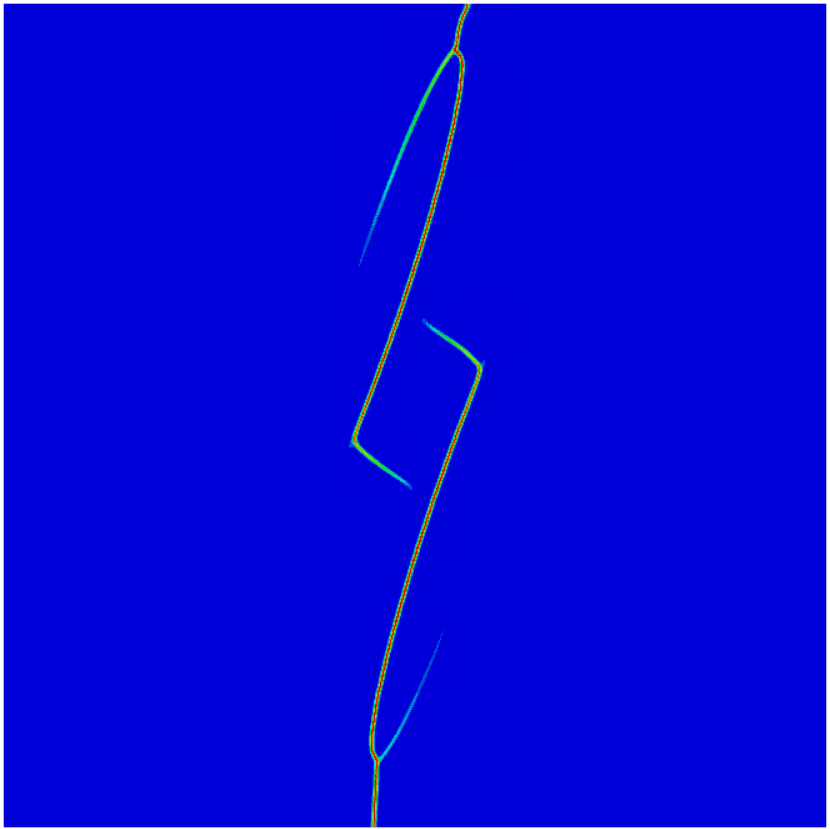} 
\caption{Reference}
\label{ref_void}
\end{subfigure}
  \begin{subfigure}{0.222\textwidth}
    \includegraphics[width=\linewidth]{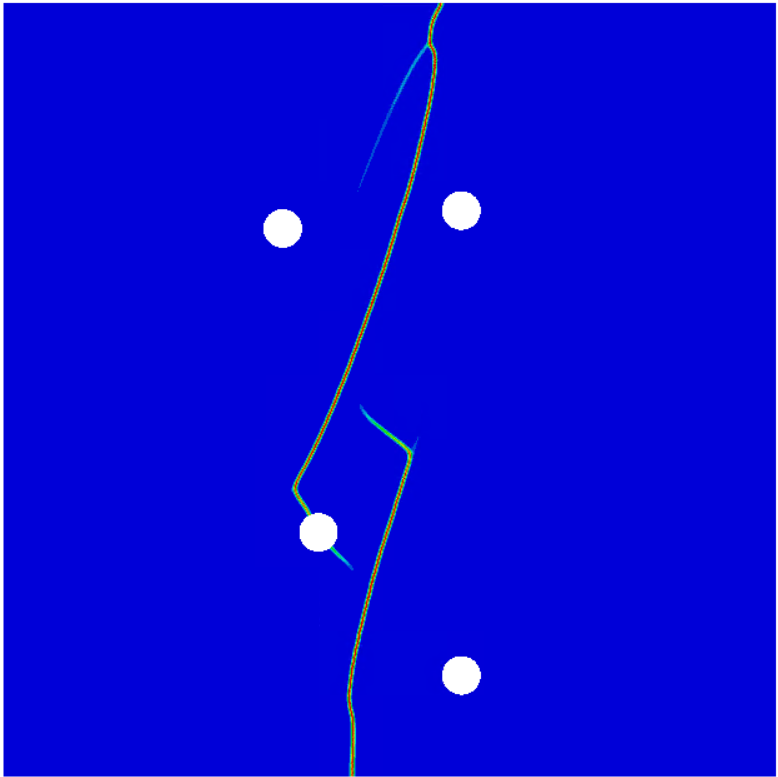} 
    \caption{void 1}
    \label{void1}
  \end{subfigure}
  \hspace{0.5em}  % Adjusted spacing
  % Fifth picture (Placeholder)
  \begin{subfigure}{0.222\textwidth}
    \includegraphics[width=\linewidth]{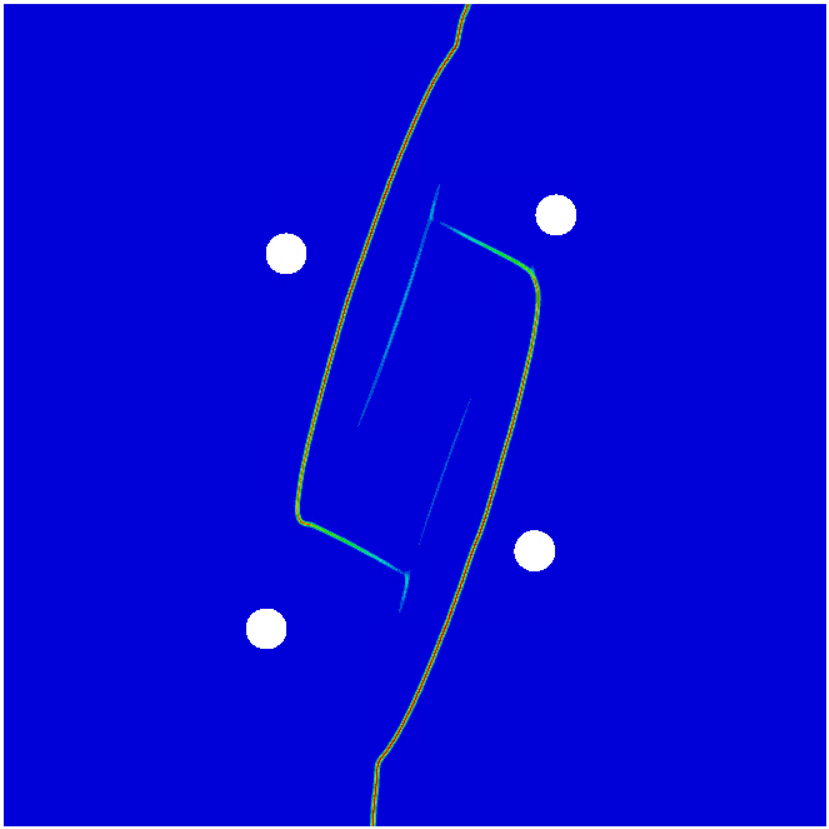} 
    \caption{void 2}
    \label{void2}
  \end{subfigure}
  \hspace{0.5em}  % Adjusted spacing
  % Sixth picture (Placeholder)
  \begin{subfigure}{0.222\textwidth}
    \includegraphics[width=\linewidth]{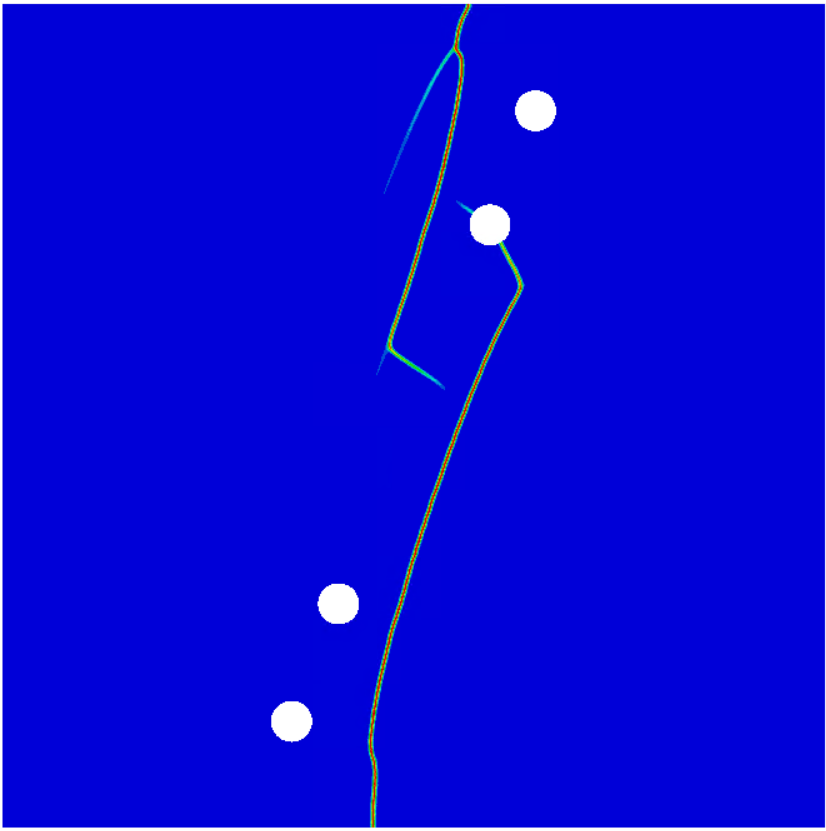} 
    \caption{void 3}
    \label{f}
  \end{subfigure}       
  \caption{Effect of voids on the final fracture paths.}
  \label{void3}
\end{figure}
\begin{figure}[H]
\center
    \includegraphics[width=0.6\textwidth]{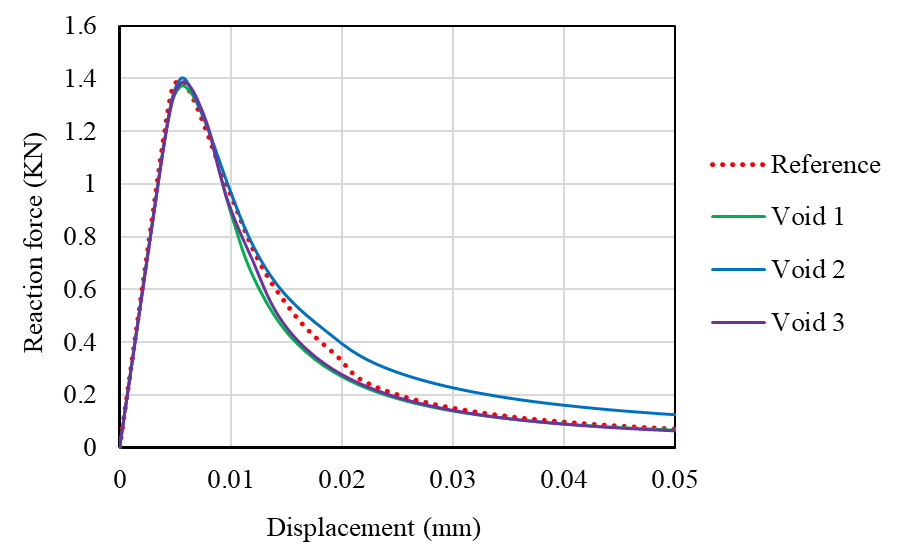}
    \caption{Force-displacement curves derived from FE simulations using CPFM approach. While all material attributes are kept constant, only the spatial distribution of voids are varied.}
    \label{voids}
\end{figure}

In addition, the effect of porosity within the microstructure on the final fracture energy is also investigated. Four circular voids with a radius of 2.5 cm are distributed within the microstructure in three different patterns and the resulting fracture paths and fracture energy are compared with the reference case without voids. The reference microstructure is shown in Figure~\ref{type4}. It can be seen that despite the removal of almost 3\% of the total volume fraction of the microstructure by incorporating voids, the fracture energy in Figure~\ref{void2} is higher than in the original case. This is because the voids have indirectly influenced the fracture paths, resulting in longer fracture paths in the case of Figure~\ref{void2}. 

%#########################################################################
\section{Conclusion}\label{sec5}
This study analyses four numerical models for simulating crack initiation and growth in heterogeneous materials with an interface phase, focusing on their implementation complexity and computational cost. The methods include the standard phase-field fracture, the cohesive phase-field fracture, the cohesive zone model and a hybrid model where the cohesive zone handles interface debonding and the cohesive phase-field simulates matrix cracking. The conclusions are based on a 2D single inclusion model, considering that the interface thickness is comparable to other phases. The main findings are:
\begin{itemize}
    \item The SPFM model does not capture cohesive behaviour in semi-brittle materials in the softening zone, and its length scale parameter \(l_{c}\) is treated as an intrinsic material property, limiting its use in small-scale simulations. The CPFM model is more suitable for multiphase materials, especially when the interface thickness is comparable to other phases. In CPFM, the length scale parameter acts as a numerical property, making it ideal for small-scale simulations. It is also less computationally demanding and easier to implement than the hybrid approach.
    \item Composite design with targeted properties can be achieved by adjusting inclusion arrangement, modifying interface properties, adding voids or a combination of these factors. This study shows that by strategically arranging inclusions or interface properties, fracture toughness can be significantly improved while keeping all other parameters constant.
\end{itemize}
Moreover, if the interface is much thinner than other phases, the hybrid model is the best choice for fracture simulation because thin interfaces require a fine mesh in order to be captured appropriately in FE simulations. Another option is to artificially thicken the interface and adjust its properties so that the CPFM model can be used. This makes the CPFM model less sensitive to interface thickness, making it a good option for multiphase materials with isotropic phases. See \cite{zhou2022interface} for more information.
\subsection{Future development}
Among the fracture models analysed, the CPFM approach is the preferred method for the simulation of heterogeneous multiphase solids where the thickness of the interface is comparable to that of other phases. Future work could focus on the optimisation of microstructures to improve fracture properties and the use of data-driven surrogate modelling techniques to predict mechanical behaviour and fracture paths using the CPFM model for generating training data. Further development could include neural network architectures to scale interfacial properties, allowing the transition from the hybrid model to the CPFM model. It is important to note that many other fracture simulation methods for heterogeneous materials have not been compared in this study. Two methods worth mentioning are finite cohesive zone modelling and specific PFM methods that model interface softening using PFM approaches. Future research could include a more comprehensive comparison of these advanced models that address the limitations of the classical CZM and standard phase-field models used here, potentially leading to more accurate and effective simulations.  
%%%%%%%%%%%%%%%%%%%%%%%%%%%%%%%%%%%%%%%%%%
\vspace{6pt} 
\newpage
%%%%%%%%%%%%%%%%%%%%%%%%%%%%%%%%%%%%%%%%%%
\authorcontributions{Rasoul Najafi: Conceptualization, methodology, software, validation, visualization, investigation, resources, writing original draft, review, and editing. Natalie Rauter: Supervision, review and editing, project administration, funding acquisition. Rolf Lammering: Supervision, review and editing, project administration, funding acquisition. Richard Ostwald: Review and editing, Shahed Rezaei: Review and editing}
\funding{This research is funded by dtec.bw - Digitalization and Technology Research Center of the Bundeswehr, which we gratefully acknowledge. dtec.bw is funded by the European Union – NextGenerationEU. The author Shahed Rezaei would like to thank the Deutsche Forschungsgemeinschaft (DFG) for the funding support provided to develop the present work in the project Cluster of Excellence “Internet of Production” (project: 390621612).}

\dataavailability{The codes and data associated with this research are available upon request and will be published online following the official publication of the work.} 
%\acknowledgments{}
\conflictsofinterest{The authors declare no conflict of interest} 
%%%%%%%%%%%%%%%%%%%%%%%%%%%%%%%
%% Optional
\appendixtitles{no} % Leave argument "no" if all appendix headings stay EMPTY (then no dot is printed after "Appendix A"). If the appendix sections contain a heading then change the argument to "yes".
\appendixstart
\appendix
\section[\appendixname~\thesection]{Phase-Field Fracture Models}\label{Appendix_A}
\subsection[\appendixname~\thesubsection]{Cohesive Phase-Field Fracture Model}
The Cohesive Phase-Field Model (CPFM) releases fracture energy progressively compared to the instantaneous dissipation in standard models. The CPFM is effective for quasi-brittle materials and is insensitive to the length scale parameter \( l_c \).

The damage degradation function in CPFM is defined as
\begin{linenomath}
\begin{equation}
    \omega(\phi) = \frac{(1-\phi)^2}{(1-\phi)^2 + a_{1} \phi + a_{1} a_{2} \phi^2 + a_{1} a_{2} a_{3} \phi^3},  
\end{equation}
\end{linenomath}
with coefficients \(a_1\), \(a_2\), and \(a_3\) given by

\begin{linenomath}
\begin{equation}
  \begin{cases}
    \displaystyle a_{1} = \frac{2 E G_{c}}{\sigma_{u}^2 \pi l_{c}}, \\[1.5ex]
    \displaystyle a_{2} = 2 \left( -2 k_{0} \frac{G_{c}}{\sigma_{u}^2} \right)^{\frac{2}{3}} - \left(p + \frac{1}{2}\right), \\[1.5ex]
    \displaystyle a_{3} = 
    \begin{cases}
      0 & \text{if} \quad p > 0, \\[1.5ex]
      \frac{1}{a_{2}} \left[ \frac{1}{8} \left( \frac{d_{u} \sigma_{u}}{G_{c}} \right)^2 - (1 + a_{2}) \right] & \text{if} \quad p = 0,
    \end{cases} 
  \end{cases}
\end{equation}
\end{linenomath}
where \( E \) is the elasticity modulus, \( p \) is an exponent related to material properties, and \( k_0 \) and \( d_u \) represent the initial slope and ultimate crack opening, respectively.

The CPFM crack geometric function is defined as
\begin{linenomath}
\begin{equation}
    \alpha(\phi) = 2 \phi - \phi^2,
\end{equation}
\end{linenomath}
resulting in \( c_0 = \pi \).

The history variable is defined as
\begin{linenomath}
\begin{equation}
  \begin{cases}
    H(\boldsymbol{x},t) = \max \left\{ \psi_{\mathrm{eq}}^0, \max \psi_{\mathrm{eq}}(\boldsymbol{\varepsilon}(x, t)) \right\}, \\[1.5ex]
    \psi_{\mathrm{eq}}^0 = \frac{1}{2E} \sigma_{u}^2,
  \end{cases}
\end{equation}
\end{linenomath}
where \( \psi_{\mathrm{eq}}^0 \) is a threshold for damage initiation.

\subsection[\appendixname~\thesubsection]{Standard Phase-Field Fracture Model}
The AT1 and AT2 models represent distinct formulations for simulating material fracture in the Standard Phase-Field Model (SPFM). The damage degradation function in SPFM is expressed as
\begin{linenomath}
\begin{equation}
    \omega(\phi) = (1 - \phi)^2 + k,
\end{equation}
\end{linenomath}
where \( k \) is a small numerical parameter.

The crack geometric function is defined as:
\begin{linenomath}
\begin{equation}
  \begin{cases}
    \alpha(\phi) = \phi & \text{(AT1 model)}, \\[1.5ex]
    \alpha(\phi) = \phi^2 & \text{(AT2 model)}.
  \end{cases}
\end{equation}
\end{linenomath}

The parameter \( c_{0} \) is given by
\begin{linenomath}
\begin{equation}
    c_{0} = 4 \int_{0}^{1} \sqrt{\alpha(x)} \, \mathrm{dx},
\end{equation}
\end{linenomath}
resulting in \( c_{0} = \frac{8}{3} \) for the AT1 model and \( c_{0} = 2 \) for the AT2 model.
\begin{adjustwidth}{-\extralength}{0cm}
\reftitle{References}

\PublishersNote{}
\end{adjustwidth}
\end{document}